\documentclass[10pt,twocolumn,letterpaper]{article}
\usepackage[letterpaper,hmargin=.75in,vmargin=.75in]{geometry}
\usepackage[10pt]{sigmin}
\usepackage[noadjust]{cite}
\usepackage{amsmath,amssymb,times,mathptmx,verbatim,graphicx}
\usepackage{subfigure}
\usepackage{url}

\newcommand{\beq}{\begin{equation}}
\newcommand{\eeq}{\end{equation}} 
\def\bearn{\begin{eqnarray*}}
\def\eearn{\end{eqnarray*}}
\def\barr{\begin{array}}
\def\earr{\end{array}}
  
\def\bt{BitTorrent }
\def\btns{BitTorrent} 
\def\p2p{peer-to-peer}

\begin{document}
\title{\sffamily\textbf{Clustering and Sharing Incentives in BitTorrent Systems}\vspace{0.2cm}}

\author{\sffamily\small\begin{tabular*}{14cm}{c@{\hspace{20mm}}ccc}
\normalsize{Arnaud Legout} 
	& \normalsize{\hspace{15mm}Nikitas Liogkas}
	& \normalsize{Eddie Kohler}
	& \normalsize{Lixia Zhang} \\
\noalign{\vskip.1ex}
I.N.R.I.A. & \multicolumn{3}{c}{\hspace{15mm}University of California, Los Angeles} \\
Sophia Antipolis, France & \multicolumn{3}{c}{\hspace{15mm}Los Angeles, CA, USA} \\
arnaud.legout@sophia.inria.fr &
\multicolumn{3}{c}{\hspace{15mm}\{nikitas,kohler,lixia\}@cs.ucla.edu}
\vspace{0.2cm}
\end{tabular*}}

\date{}
\toappear{Technical report\\ inria-00112066, version 1 - 21 November
  2006}

\maketitle
\sloppy

\begin{abstract}
Peer-to-peer protocols play an increasingly instrumental role in 
Internet content distribution. Consequently, it is important to 
gain a full understanding of how these protocols behave in practice and how 
their parameters impact  overall performance. 
We present the first experimental investigation of the peer selection
strategy of the popular BitTorrent protocol in an 
instrumented private torrent.
By observing the decisions of more than 40 nodes, we validate
three BitTorrent properties that, though widely believed to hold, have
not been demonstrated experimentally. These include the clustering of
similar-bandwidth peers, the effectiveness of BitTorrent's sharing
incentives, and the peers' high average upload
utilization.
In addition, our results show that BitTorrent's new choking algorithm
in seed state provides uniform service to all peers, and that an
underprovisioned initial seed leads to
the absence of peer clustering and less effective sharing
incentives.
Based on our observations, we provide guidelines for seed provisioning 
by content providers, and discuss a tracker protocol extension that addresses
an identified limitation of the protocol.

\end{abstract}

\section{Introduction}
In just a few years, peer-to-peer (P2P) content distribution has managed to
enter the class of applications generating a significant amount of Internet 
traffic~\cite{karagiannis04}. 
This widespread adoption of P2P protocols for delivering large data volumes
to geographically dispersed peers is arguably due to their scalability
and robustness properties.
Understanding the mechanisms that affect the performance of such 
protocols, and designing improved algorithms to overcome existing
shortcomings, is critical to the continued success of P2P data delivery. 
This paper presents a detailed study of BitTorrent, one of 
the most popular P2P content distribution protocols. We measure BitTorrent's 
performance in a controlled environment, running real experiments on a 
private testbed for a variety of scenarios. 

There have recently been several attempts to analyze BitTorrent system behavior, 
as well as experimentally evaluate its fundamental algorithms.
Some researchers have formulated analytical models for the problem of efficient 
data exchange among peers. For example, Yang \textit{et al.}~\cite{yang04} 
study the service capacity of BitTorrent-like protocols. They show that it 
increases exponentially at the beginning of the download session, and scales 
well with the number of participating peers. 
In addition, measurement studies of actual download traces have attempted to
shed more light into the success of the protocol.
For example, Pouwelse \textit{et al.}~\cite{pouwelse05} study the file 
popularity, file availability, and content lifetime of numerous download sessions.

However, certain properties of previous studies prevented them from
accurately evaluating the dynamics of \bt algorithms and their impact on the
overall performance.  The analytical models provide valuable insight,
but typically make unrealistic assumptions to simplify analysis, such
as giving all participants global system knowledge~\cite{yang04,
  qiu04}; actual download traces can differ substantially from their
predictions~\cite{guo05, pouwelse05}.  Previous measurement studies
have evaluated peers connected to public
torrents~\cite{izal04,guo05,pouwelse05}.  These studies provide useful
information about the behavior of deployed BitTorrent systems, but the
information available from a public torrent is coarse-grained, and
does not explain individual peer decisions during the download.  A
more recent study does evaluate those decisions, but only from the
viewpoint of a single peer~\cite{legout06}.

In order to overcome these limitations, we evaluate the performance of 
BitTorrent by running extensive experiments in a controlled environment.
In particular, we focus on the so-called 
\emph{choking algorithm} for peer selection, which is arguably the driving
factor behind the protocol's high performance~\cite{cohen03}. 
This approach allows us to examine the behavior of BitTorrent systems under a 
microscope, and evaluate the impact of different parameters on system performance.
In the process, we validate certain properties of the choking algorithm that 
are widely believed to hold, but have not been demonstrated experimentally. 
In addition, we identify new properties and offer insights into the behavior of 
the choking algorithm in different scenarios, as well as into the impact of proper
provisioning of the initial seed on performance.

The contributions of this work are three-fold. First, we demonstrate that the choking algorithm 
enables good clustering of similar-bandwidth peers, ensures effective
sharing incentives by rewarding
peers who contribute with high download rates, and achieves high upload utilization for 
the majority of the download duration. These properties have been hinted at in previous work;
this study constitutes their first experimental validation.
Second, we pinpoint newly observed properties and limitations of the
choking algorithm. 
The new choking algorithm in seed state provides service to all peers
uniformly.  As a result, if the seed is underprovisioned, clustering is poor
and peers tend to finish their downloads at
the same time, independently of how much they contribute.
Finally, based on our observations, we provide guidelines for seed provisioning 
by content providers, and discuss a tracker protocol extension that addresses
an identified limitation of the protocol, namely the low upload utilization at the
beginning of a torrent's lifetime.

The rest of this paper is organized as follows.
Section~\ref{background} provides a brief description of the
BitTorrent protocol and an explanation of the choking
algorithm, as implemented in the official BitTorrent client.  
Section~\ref{methodology} describes our methodology and the rationale
behind our experiments, while
Section~\ref{results} presents our experimental results.
Section~\ref{discussion} discusses our proposed seed
provisioning guidelines, and the proposed tracker protocol extension. Lastly,
Section~\ref{related} sets this study in the context of related work,
and Section~\ref{conclusion} concludes.

\section{Background}
\label{background}
BitTorrent is a peer-to-peer content distribution protocol that has been
shown to scale with the number of participating peers. In particular, a \bt system
capitalizes on the upload capacity of each peer in order to increase
the global system capacity as the number of peers increases. 
A major factor behind \btns's success is its built-in incentive mechanism, as enforced by the
choking algorithm, which is intended to motivate peers to contribute
data. The rest of this section introduces the terminology used in this
paper, describes \btns's operation in detail, and focuses 
on the choking algorithm in particular.

\subsection{Terminology}
\label{sec:terminology}
The terminology used in the \bt community 
is not standardized. For the sake of
clarity, we define here the terms used throughout this
paper.

\begin{itemize}
\item
\textbf{Pieces and Blocks} 
Content transferred using \bt is split into
  \textit{pieces}, and each piece is split into multiple \textit{blocks}. Blocks
  are the transmission unit in the network, but peers can only
  share complete pieces with others.

\item
\textbf{Interested and Choked} 
We say that peer $A$ is
  \textit{interested} in peer $B$, when $B$ has pieces that 
  $A$ does not have. Conversely, peer $A$ is \textit{not interested}
  in peer $B$, when $B$ only has a subset of the pieces of 
  $A$. We also say that peer $A$ is being \textit{choked} by peer $B$, when $B$
  has decided not to send any data to $A$. Conversely, peer $A$
  is being \textit{unchoked} by peer $B$, when $B$ is willing to send data to
  $A$. 
  Note that this does not necessarily mean that peer $B$ is uploading data to 
  peer $A$. It just means that $B$ \emph{is willing} to upload to
  $A$, whenever $A$ makes a piece request to $B$.

\item
\textbf{Peer Set} 
Each peer maintains a list of other peers, to which it has open TCP connections.
  We call this list the \textit{peer set}. This 
  is also known as the neighbor set.

\item 
\textbf{Local and Remote Peers} When we illustrate the choking algorithm below
 we take the point of view of a single peer that we call
 \textit{local peer}. We refer to peers that are in
 the local peer's peer set as \textit{remote peers}.

\item
\textbf{Leecher and Seed} 
A peer can be in one of two states: the leecher
    state, when it is still downloading pieces of the content,
  and the seed state when it has all the pieces
  and is sharing them with others. In short, we say that a peer is a
  \textit{leecher} when it is in the leecher state, and a \textit{seed}
  when it is in the seed state.

\item
\textbf{Initial Seed} 
The \textit{initial seed} is the peer that initially offers the content for download.
There can be more than one initial seed. In this paper, we consider
only the case of a single initial seed. 

\item
\textbf{Rarest-First Algorithm} 
The \textit{rarest-first algorithm} is the piece
  selection strategy used in \btns, also known as the local rarest-first algorithm, 
  since it bases its decisions on limited local knowledge of the torrent.
  Each peer maintains a list of the number of copies of each piece that peers in its peer 
  set have. It uses this information to define a \textit{rarest pieces set}, 
  which contains the indices of all the pieces with the least number of copies.
  This set is updated every time a remote peer in the peer set acquires
  a new piece, and is consulted for the selection of the next piece to download.

\item
\textbf{Choking Algorithm} 
The \textit{choking algorithm} is the peer selection
  strategy used in \btns, also known as the tit-for-tat algorithm. We provide a detailed description of this
  algorithm in section~\ref{choking_algorithm}.  

\item \textbf{Official \bt Client} The official \bt client
  \cite{btsite}, also known as \textit{mainline} client, was initially
  developed by Bram Cohen and is now maintained by the company he
  founded.
\end{itemize}

\subsection{BitTorrent Operation}
A \textit{torrent} is a set of peers cooperating to download the same
content using the \bt protocol.  Prior to distribution,
the content is
divided into multiple pieces, and each piece into multiple blocks.  A
\textit{metainfo file}, also called a torrent file, containing all
information necessary for the download process is created. It includes
the number of pieces, SHA-1 hashes for all the pieces, and the IP
address and port number of the so-called \textit{tracker}. The 
hashes are used by peers to verify that a piece has been received
correctly. The tracker is the only centralized component of the system.
It is not involved in the actual distribution of the content, but rather, it keeps
track of all peers currently participating in the download and also collects
statistics for all peers.
In order to join a torrent, a peer retrieves the metainfo file out
of band, usually from a well-known website. It then contacts the
tracker that responds with a peer set of randomly selected peers,
which might include both seeds and leechers.  The newly arrived peer
starts contacting peers in this set, requesting different pieces.

Most clients nowadays implement the rarest-first algorithm for
piece selection. 
According to that, peers select the next piece to
download from their rarest pieces set. They are able to determine which
pieces other peers have based on a \textit{bitfield} message exchanged upon
new connections, which contains the list of all pieces
a peer has. Peers also send \textit{have} messages
when they successfully receive and verify a new piece. These messages are typically sent to
all peers in their peer set.

The selection that determines which peers to exchange data
with is made via the so-called choking algorithm. This algorithm gives
preference to those peers who upload data at high
rates.  Once per \textit{rechoke period}, typically every ten seconds,
each peer reconsiders the receiving data rates from all
the peers in its peer set. It then selects the fastest ones and
uploads only to those for the duration of the
period. 
In BitTorrent parlance, a peer unchokes the fastest uploaders via a \textit{regular unchoke},
and chokes all the rest.  Furthermore, an additional peer is
randomly unchoked once every third rechoke period, by means of an
\textit{optimistic unchoke}.  

Seeds, who do not need to download any
pieces, have to follow a different strategy.  Most implementations
dictate that seeds unchoke those leechers that download content at the
highest rates, in order to better utilize the available seed upload
capacity.  The official BitTorrent client~\cite{btsite}, however,
starting with version 4.0.0, has introduced an entirely new algorithm
in seed state. In this paper, we perform the first detailed experimental evaluation
of this algorithm and show that it contributes to an even more efficient
utilization of the seed's bandwidth.

\subsection{Choking Algorithm}
\label{choking_algorithm}
We now describe the choking algorithm in detail, as implemented in the
official client, version
4.0.2. This algorithm was introduced to guarantee a high level of data
exchange reciprocation, and is one of the main factors behind
BitTorrent's sharing incentives: peers that do not contribute should
not be able to attain high download rates, since such peers will be
choked by others.  As a consequence, \textit{free-riders}, i.e., peers
that never upload, should be penalized.
The algorithm does not prevent all free-riding~\cite{liogkas06, locher06},
but we show it performs well in a variety of circumstances.

The choking algorithm is different for leechers and seeds. In leecher state, 
a fixed number of remote peers are unchoked every rechoke period. This number of 
parallel uploads is determined by the imposed limit on upload bandwidth, unless 
specified explicitly by the user. For example, for an upload limit greater than or 
equal to 15 kB/s but less than 42 kB/s this number is four.
In the following, we assume that the number of parallel uploads is set to $n$. 

In leecher state, the choking algorithm is executed periodically at every rechoke
period, i.e., every ten seconds
and, in addition, whenever an unchoked and interested peer leaves the peer set, or whenever 
an unchoked peer switches its interest state.  As a consequence, the
time interval between two executions of the algorithm can be much shorter
than the duration of the rechoke period. 
Every time the choking algorithm is executed, we say that a new \textit{round} starts, 
and the following steps are taken.
\begin{enumerate}
\item \label{step:l1} Interested leechers are ordered according to 
  their observed upload rates to the local peer. 
  However, the local peer ignores leechers that have not sent it any data in the
  last 30 seconds.  These \emph{snubbed}
  peers are excluded from consideration in order to guarantee that only contributing peers are
  unchoked.
\item \label{step:l2} The $n-1$ fastest of these leechers are unchoked via a so-called 
  \emph{regular unchoke}.
\item \label{step:l3} In addition, a candidate peer is chosen at random to be unchoked via
  a so-called \emph{optimistic unchoke}. 
  \begin{enumerate}
  \item \label{step:l4} If the candidate peer is interested in the local peer,
    it is indeed unchoked via an optimistic unchoke and 
    the round is completed.
  \item \label{step:l5} Otherwise, the candidate peer is unchoked anyway, but the
  algorithm repeats step~\ref{step:l4} with a new randomly-chosen candidate.
  \end{enumerate}
  The round completes when an interested
  peer is found or when there are no more peers, whichever comes first.
\end{enumerate}

Although more than $n$ peers can be unchoked by the algorithm, 
only $n$ interested peers can be unchoked in the same round. 
Unchoking uninterested peers improves reaction time in case one of those peers 
becomes interested during the following rechoke period: data transfer can begin 
right away without waiting for the choking algorithm.
Optimistic unchokes serve two major purposes. 
They allow continuous evaluation of the upload contributions of all peers in the peer set, 
in an effort to discover better partners. They also enable new peers that do not have any 
pieces yet to bootstrap into the torrent by giving them some first
pieces without requiring reciprocation.

For the seed state, older versions of the official client, as well as many current
versions of other clients, performed the same steps as in the leecher
state with the only difference
that the ordering performed in step~\ref{step:l1} was based on observed
download rates from the seed, rather than upload rates.
Consequently, peers with high download capacity were favored independently of their 
contribution to the torrent, a fact that could be exploited by free-riders~\cite{liogkas06}.
Starting with version 4.0.0, the official client introduced an
entirely new choking algorithm 
in seed state.  We are not aware of any other documentation of this new algorithm, nor of 
any other implementation that uses it. According to this algorithm, the same fixed 
number of $n$ parallel uploads as in the leecher state is performed
during every rechoke period. However, the peer selection criteria are
now different. 

The algorithm is executed periodically at every rechoke period, i.e., every ten seconds,
and, in addition, whenever an unchoked and interested peer leaves the peer set, or whenever 
an unchoked peer switches its interest state.
Every time the choking algorithm is executed, a new round starts, 
and the following steps are taken.

\begin{enumerate}
\item \label{step:s1} The leechers that are interested and unchoked
  are ordered according to the time they were last unchoked (most recently
  unchoked peers first). This step only considers leechers that were unchoked recently
  (less than 20 seconds ago) or that have pending requests for blocks (to ensure that
  they get the requested data as soon as possible).
  In case of a tie, leechers are ordered according to their download rates 
  from the seed, fastest ones first.
  Note that as leechers are not expected to upload anything to seeds, the notion
  of snubbed peers does not exist in seed state.

\item \label{step:s2} The number of optimistic unchokes to perform \emph{over the 
  duration of the next three rechoke periods}, i.e., 30 seconds, is determined using a heuristic.  
  These optimistic unchokes are uniformly spread over this duration, performing $n_o$ 
  optimistic unchokes per rechoke period. 
  Due to rounding issues, $n_o$ can be different for each of the three rechoke periods.
  For instance, when the number of parallel uploads is 4, the heuristic dictates that only 2 
  optimistic unchokes must be performed in the entire 30-second period. Thus, 1 optimistic unchoke 
  is performed during each of the first two rechoke periods and none during the last.
\item \label{step:s3} The first $n-n_o$ leechers in the ordered list calculated in 
  step~\ref{step:s1} are unchoked via regular unchokes.
\end{enumerate}

Step \ref{step:s1} is the key of the new algorithm in
seed state. Leechers are no longer unchoked based on their 
download rates from the seed, but mainly based on the time of their last unchoke. 
According to the official client's version notes, this new choking
algorithm in seed state aims at reducing the amount of duplicate data a seed needs to upload before it has pushed
out a full copy of the content into the torrent. 

Some other clients have implemented a \emph{super-seeding} 
feature with similar goals, in particular to assist a 
service provider with limited upload capacity in seeding a large torrent.
A seed in super-seeding mode
masquerades as a normal leecher with no data. 
As other peers connect to it, it will advertise a piece that it has never uploaded 
or that is very rare. After uploading this piece to a leecher, the seed will 
not advertise any new pieces to that leecher until it sees another peer's
advertisement for the piece, indicating that the leecher has indeed shared 
the piece with others.
This algorithm has anecdotally resulted in much higher seeding 
efficiencies by reducing the amount of redundant pieces uploaded by the seed, and limiting 
the amount of data sent to peers who do not contribute~\cite{btwikispec}. A single seed running in this mode
is supposed to be able to upload a full copy of the entire content after only 
uploading 105\% of the content data volume. 
Since the official client has not implemented this super-seeding feature, our experiments
do not measure its effect on the efficiency
of the initial seed.
Instead, we measure the number of duplicate pieces uploaded by the
initial seed when employing the new
choking algorithm in seed state.

\section{Methodology}
\label{methodology}

\subsection{Experimental Setup}
All our experiments were performed with private torrents on the 
PlanetLab experimental platform~\cite{planetlab}. 
PlanetLab's 
convenient tools for collecting measurements from geographically 
dispersed clients greatly facilitated our work.
For instance, in order to deploy and launch BitTorrent clients 
on the PlanetLab nodes, we utilize the \textit{pssh} tools~\cite{pssh}.
PlanetLab nodes are typically not behind NATs, and they keep all their ports open,
so each peer in our experiments can be uniquely identified by its IP address. 
We consciously chose to experiment on private torrents
in order to examine both per-peer decisions and the resulting overall
torrent behavior. 
Private torrents allowed us to observe and record the
behavior of all peers throughout the torrent's lifetime.
It also let us vary experimental parameters, such as upload bandwidth
limits of the leechers and the seed. 
This in turn helped us identify conditions that improve or hinder overall
performance and distinguish which factors are responsible for observed
behavior.

We performed experiments on different torrent configurations, and
repeated each experiment run several times. During each experiment, leechers
download a single 113 MB file that consists of 453 pieces, 256 kB each.

PlanetLab's available bandwidth is unusually high for typical torrents;
we enforce upload limits on the leechers and seed to model realistic scenarios.
However, we do not impose any download limits whatsoever, nor do we attempt
to match our upload limits to inherent limitations of PlanetLab nodes.
Thus, for example, we might end up imposing a high upload limit on a 
node that cannot possibly send data that fast, due to network or other problems.

We perform our experiments with a single initial seed, and in all experiments, 
all leechers join at the same time, simulating a flash 
crowd scenario. Although the behavior of a torrent might be different
with other peer arrival patterns, we are interested in examining peer behavior
under circumstances of high load.
The initial seed stays connected to the torrent for the duration of
each experiment, while
leechers disconnect from the torrent immediately after completing their download.

We collect our measurements by utilizing a modified version of the
official BitTorrent implementation~\cite{btsite}, 
which we instrumented to record interesting events and peer interactions.
The instrumented client is based on version 4.0.2 of the official client, 
which was released in May 2005. 
Our client is publicly available for download~\cite{btinstrumented}.
The instrumentation we collect consists of a log of each message sent or 
received along with the content of the message, a log of each state change,
the rate estimates used by the choking algorithm, and a log of other  
information, such as internal states of the choking algorithm.

\subsection{Torrent Configurations}
We experimented with several torrent configurations.
The parameters we changed from configuration to configuration are the
upload limits for the seed and leechers, and the upload bandwidth
distribution of leechers.
As mentioned before, leecher download bandwidth is never 
artificially limited, although in some cases, local network characteristics may impose 
an effective upload or download limit.
Since any leecher could potentially download as fast as any other,
differences in observed download rates originate solely in BitTorrent's
choking algorithm.

We ran experiments with the following configurations:
\begin{itemize}
\item \emph{Two-class}: leechers are divided into two categories with
  different imposed upload limits.  This configuration enables us to
  observe system behavior in highly bipolar scenarios.  Our
  experiments involve similar numbers of slow peers, with 20~kB/s
  upload limit, and fast peers, with 200~kB/s upload limit.

\item \emph{Three-class}: leechers are divided into three categories
  with different imposed upload limits.  This configuration helps us
  in identifying the qualitative behavioral differences of more
  distinct classes of peers.  Our experiments involve similar numbers
  of \emph{slow} peers, with 20~kB/s upload limit; \emph{medium}
  peers, with 50~kB/s upload limit; and \emph{fast} peers, with
  200~kB/s upload limit.

\item \emph{Uniform}: upload limits are imposed on leechers according
  to a uniform distribution, with a small 5 kB/s step. In our
  experiments the slowest leecher has an upload limit of 20 kB/s, the
  second slowest a limit of 25 kB/s, and so on.  This configuration
  provides insight into the behavior of more homogeneous torrents.
\end{itemize}

Our graphs in the next section correspond to experiments run with the
three-class configuration, but the conclusions we draw 
accord well with the results of other experiments as well. We stress
distinctions where appropriate.

In our experiments, we have considered 
both a well-provisioned and an underprovisioned initial seed.
Seed upload capacity has already been shown to be critical to
performance at the beginning of a torrent's lifetime, before the seed
has uploaded a complete copy of the content~\cite{legout06,
  bharambe06}.  However, it is not clear what the impact of an initial seed with
limited capacity is on system properties. Moreover,
the capacity threshold below which a limited initial seed adversely
impacts the system performance is not trivial to discover.  The
correct provisioning of the initial seed is fundamental for content
providers, in order for them to support torrents that support high system
capacity.  We attempt to sketch a possible answer in
Section~\ref{sec:seed-provisioning} based on our experimental results.

We also ran preliminary experiments where the initial seed disconnects
after uploading an entire copy of the content, but leechers remain
connected after they complete their download, becoming seeds for a short
period.  Peers in these experiments have somewhat lower completion times
than configurations with a single seed and immediate leecher disconnection,
but appear otherwise similar.

All our experiments are performed with collaborative peers, i.e.,
peers that never change their upload capacity during a download, or
disconnect before receiving a complete copy of the content. However, by
considering different upload capacities, and observing 
the resulting impact on the download rates of peers,
we can obtain an initial understanding of \bt system properties
in the presence of selfish peers, i.e., peers that want to maximize
their utility in the system by abusing protocol mechanisms. 

\subsection{Experiment Rationale}
\label{rationale}
The goal of our experiments is to understand the dynamics and
evaluate the efficiency of the choking algorithm. To reach this goal,
we consider in this work four metrics.
\begin{description}
\item[Clustering:] The choking algorithm aims to encourage high peer reciprocation
  by favoring peers who contribute.
  Therefore, we expect that peers 
  will more frequently unchoke other peers with similar 
  upload speeds, since those are the peers that can reciprocate with high enough rates. 
  This hypothesis has also been formulated by Qiu \textit{et al.}~\cite{qiu04} in their analytical model of \btns. 
  Consequently, we expect
  the choking algorithm to converge toward good clustering shortly
  after the beginning of the download, by grouping together peers
  with similar upload capacity. This property, however, has never
  been experimentally verified, and it is not clear whether it is always true. Indeed,
  let's consider a simple example. Peer $A$ unchokes peer $B$, because $B$
  has been uploading data at a high rate to $A$. Yet, in order for peer $B$ to continue uploading 
  to peer $A$, $A$ should also start sending data to $B$ at a high rate. The only 
  way to initiate such a reciprocative relationship is via an optimistic unchoke.
  Since optimistic unchokes are performed at random, it is not clear whether $A$ and $B$ will
  ever get a chance to interact.
  Therefore, in order to preserve clustering, optimistic unchokes
  should successfully initiate interactions between peers with similar 
  upload speeds. In addition, such interactions should persist, despite potential disruptions,
  such as optimistic unchokes by others or network bandwidth fluctuations.

\item[Sharing incentives:] A major goal of the choking algorithm is
  to give peers incentives to share data. The algorithm
  strives to prevent free-riders from monopolizing the torrent upload capacity, 
  and motivates all peers to contribute, since doing so will improve their own download rates.
  Thus, we evaluate the effectiveness of BitTorrent's sharing incentives
  by measuring how peers' upload contributions affect their download
  completion time. We expect that the more a peer contributes, the sooner it will
  complete its download. We do not expect to observe strict
  \textit{data volume fairness}, where all peers contribute the same amount of data;
  peers who upload at high rates may end up contributing much more data than others.
  However, they should be rewarded by completing their download sooner.

\item[Upload utilization:] Upload utilization constitutes a reliable
  metric of efficiency in \p2p content distribution systems, since the
  total upload capacity of all peers represents the maximum throughput
  the system can achieve as a whole. As a result, a \p2p content distribution
  protocol should aim at maximizing peers' upload utilization.
  We expect to see this high utilization in BitTorrent systems in our
  experiments. The question is how far BitTorrent is from optimal
  upload utilization levels, and which factors can adversely affect 
  utilization in specific scenarios.

\item[Seed service:] The new choking algorithm in seed state takes into account
the waiting time of peers, in addition to their observed download
rates from the seed. Thus, it should be impossible for leechers to monopolize the initial
seed, regardless of how fast they can download data. We expect to see
an even sharing of the seed upload bandwidth among all peers.
\end{description}

\section{Experimental Results}
\label{results}
We now report the results of representative experiments that demonstrate our main
observations. For conciseness, we present only results drawn from the
three-class torrent configuration, but our conclusions are consistent
with our observations from other configurations.

\subsection{Well-Provisioned Initial Seed}
\label{well_provisioned_seed}
We first examine a scenario with a \emph{well-provisioned} initial
seed, i.e., a seed that can 
sustain high upload rates. We expect this to be common for commercial torrents,
whose service providers
typically make sure there is adequate bandwidth to initially seed the
torrent.
An example might be Red Hat distributing its latest Linux distribution. 
Section~\ref{under_provisioned_seed} shows that peer behavior in the presence 
of an underprovisioned initial seed can differ substantially. 

We consider an experiment with a single seed and 40 leechers: 13 slow peers
(20~kB/s upload limit), 14 medium peers (50~kB/s upload limit), and 13 fast
peers (200~kB/s upload limit).
The seed, which is represented as peer 41 in the following figures,
is limited to upload 200~kB/s, as fast as a fast peer.  
These different peer upload limits are imposed in order to
model different levels of contribution.
The results we report are based on 13 experimental runs.  Although the
vanilla official BitTorrent implementation would set the number of
parallel uploads based on the imposed upload limit (4 for the slow, 5
for the medium, and 10 for the fast peers and the seed), we set this
number to 4 for all peers.  This ensures homogeneous conditions in the
torrent and makes it easier to interpret the results. 

\subsubsection{Clustering}
\label{clustering-fast}

\begin{figure}[t]
\centering
\includegraphics[width=2.9in]
{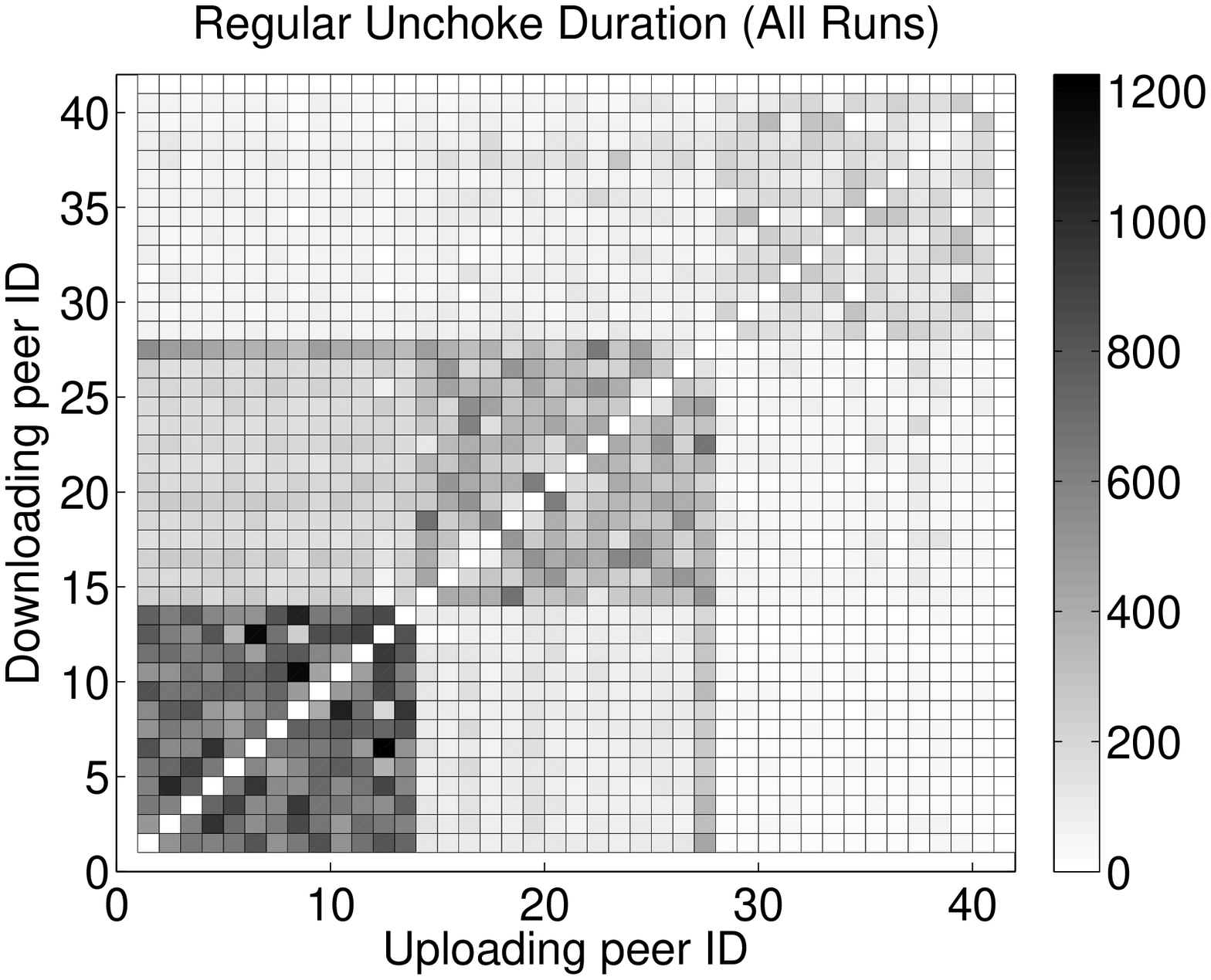}
\caption{\textmd{\textsl{Time duration that peers unchoked each other
      via a regular unchoke, averaged over all runs. Darker squares represent longer unchoke times. Peers 1 to 13 have a 20 kB/s upload limit, peers 14 to 27 have a 50 kB/s upload limit, and peers 28 to 40 have a 200 kB/s upload limit. The seed (peer 41) is limited to 200 kB/s. 
\textit{The creation of clusters is clearly visible.}}}}
\label{fig:corr-unchoke-upload-onethird-seed-200}
\end{figure}

\begin{figure}[t]
\centering
\includegraphics[width=2.9in]
{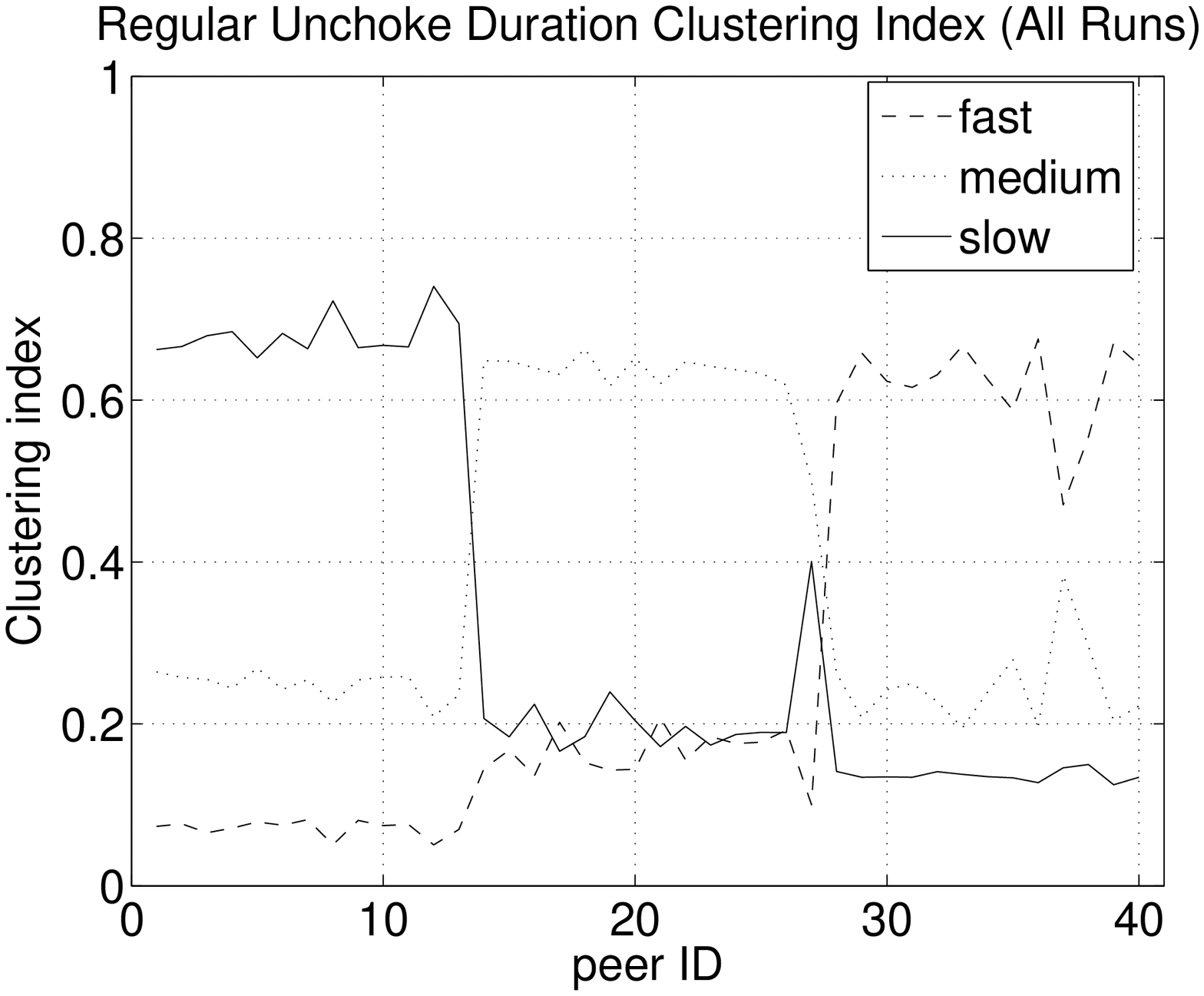}
\caption{\textmd{\textsl{Clustering indices for all peers and all runs, in the presence of a well-provisioned seed. 
Peers 1 to 13 have a 20 kB/s upload limit, peers 14 to 27 have a 50 kB/s upload limit, and peers 28 to 40 have a 200 kB/s upload limit. The seed (peer 41) is limited to 200 kB/s. \textit{Peers show a strong preference to unchoke other peers in the same class.}}}}
\label{fig:clustering-index-onethird-seed-fast}
\end{figure}

As explained in Section~\ref{rationale}, we expect to observe clustering
based on peers' upload capacities. 
Figure~\ref{fig:corr-unchoke-upload-onethird-seed-200} demonstrates that peers indeed 
form clusters. The figure plots the total time peers unchoked each 
other via a regular unchoke, averaged over all runs of the experiment. 
It is clear that peers in the same class cluster together, in the sense that they prefer
to upload to each other. This behavior becomes more apparent when considering a metric
such as the \emph{clustering index}. We define this for a given peer in 
a given class (fast, medium, or slow) as the ratio of the duration of regular unchokes to the 
peers of its class over the duration of regular unchokes to all peers. A high clustering 
index indicates a strong preference to upload to peers in the same class.
Figure~\ref{fig:clustering-index-onethird-seed-fast} demonstrates that peers in all 
classes prefer to unchoke other peers in their own class, thereby forming clusters.
Further experiments with upload limits following a uniform
distribution also show that peers have a clear preference for peers
with similar bandwidths.

\begin{figure}[t]
\centering
\includegraphics[width=2.9in]
{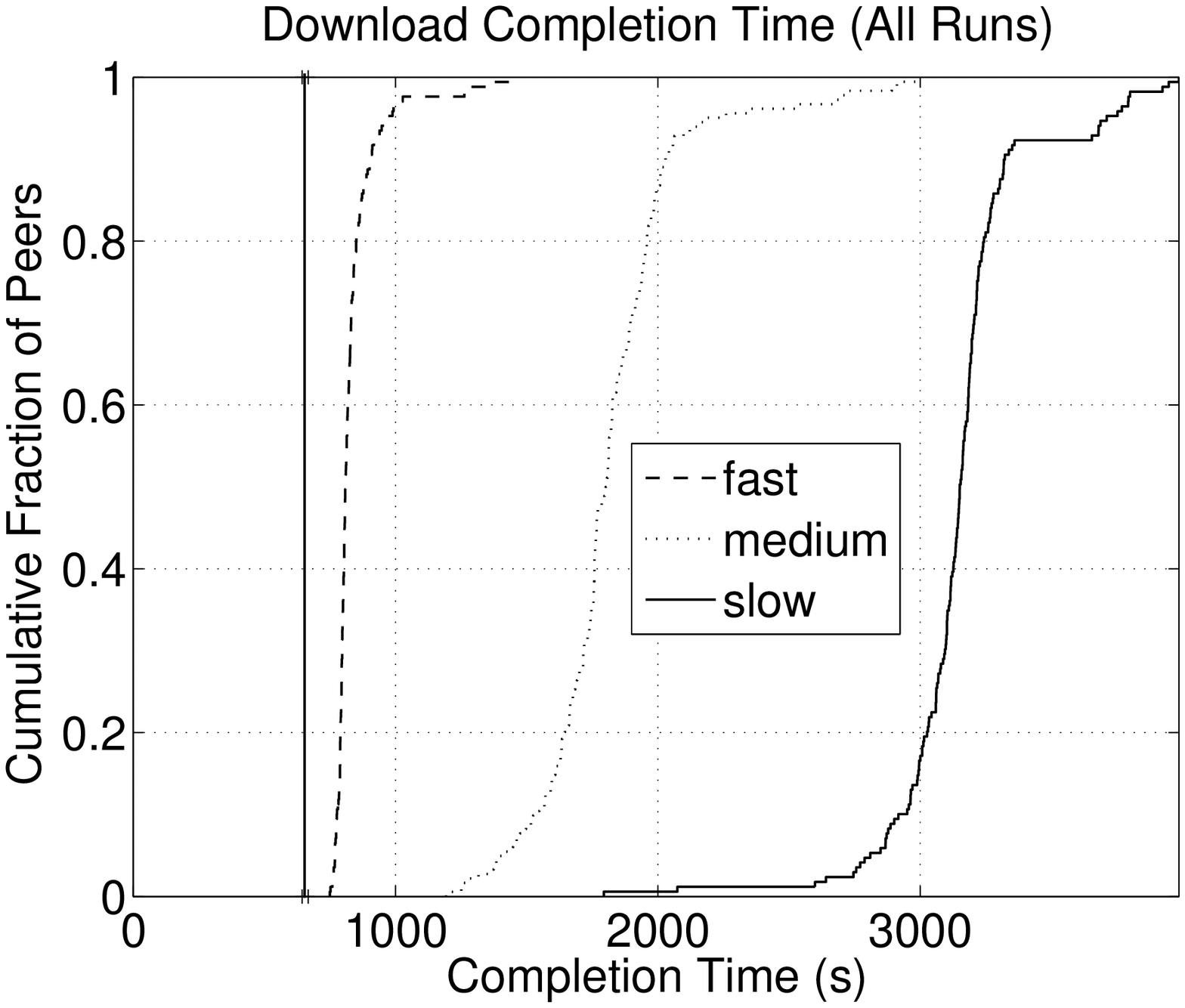}
\caption{\textmd{\textsl{Cumulative distribution of the download completion time 
for the three different classes of leechers, in the presence of a
well-provisioned seed (limited to 200 kB/s), for all runs. 
The vertical line represents the earliest possible time that the download could 
complete. \textit{Fast peers finish much earlier than slow ones.}}}}
\label{fig:completion-cdf-onethird-seed-200-class}
\end{figure}

\begin{figure}[t]
\centering
\includegraphics[width=2.9in]
{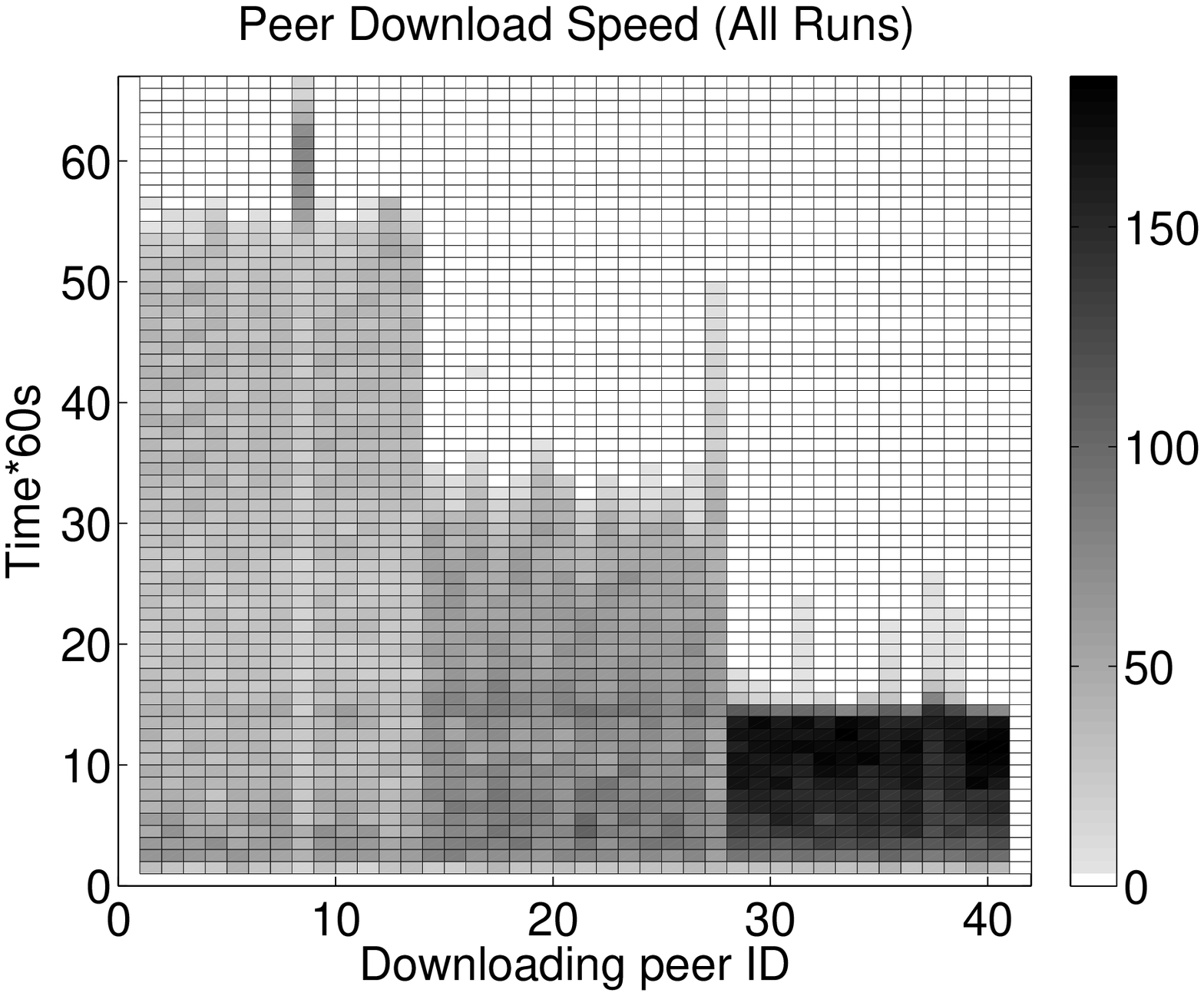}
\caption{\textmd{\textsl{Peer download speeds for all 60-sec time intervals during the 
download, averaged over all runs. Darker rectangles represent higher speeds. Peers 1 to 13 
have a 20 kB/s upload limit, peers 14 to 27 have a 50 kB/s upload limit, while peers 28 to 40 
have a 200 kB/s upload limit. The seed (peer 41) is limited to 200 kB/s. 
\textit{Peer 27 achieves lower download rates than the other peers in its class.}}}}
\label{fig:download-speed-onethird-seed-200}
\end{figure}

Although from Figure~\ref{fig:corr-unchoke-upload-onethird-seed-200} it might seem 
that slow peers show a proportionally stronger preference for their own class,
this is an artifact of the experiment. 
Slow peers take longer to complete their download (as shown in Figure  
\ref{fig:completion-cdf-onethird-seed-200-class}), and so perform a higher number of 
regular unchokes on average than fast peers.
Also notice that medium peer 27 interacts frequently with slow peers. This peer's download capacity 
is inherently limited, as seen in Figure~\ref{fig:download-speed-onethird-seed-200}
that plots observed peer download speeds over time. As a result, peer 27 stays 
connected even after all other peers of its class have completed their download. 
During that last period it has to interact with slow leechers, since those are the 
only ones left. The preference of peer 27 for slow leechers is also evident from
the spike anomaly in Figure~\ref{fig:clustering-index-onethird-seed-fast}.

Figure~\ref{fig:corr-unchoke-upload-onethird-seed-200} also shows that
reciprocation is not necessarily mutual. Slow peers frequently unchoke
medium peers, but the favor is not returned. Indeed, the slow peers
unchoked the medium peers for 501,844 seconds, as shown by
the center-left partition that is relatively dark. However, the medium
peers unchoked the slow peers for only 273,985 seconds, as shown
by the bottom-center partition that is lighter.  This lack of
reciprocation is due to the fact that slow peers are of little use to medium peers, 
since they cannot sustain high upload rates.

In summary, the choking algorithm eventually reaches an equilibrium
where peers mostly interact with others in the same class, with the occasional 
exception of optimistic unchokes, which are performed randomly. This clustering
should help keep the incentives mechanism effective.

\subsubsection{Sharing Incentives}
\label{sec:incentives-fast}
We now examine whether BitTorrent's choking algorithm provides sharing
incentives, in the sense that a peer who contributes more to the torrent
is rewarded by completing its download sooner than the rest. 
Figure~\ref{fig:completion-cdf-onethird-seed-200-class} demonstrates this to be the case. 
We plot the cumulative distribution of 
completion time for the three classes of leechers in the previous experiment. The vertical 
line in the figure
represents the \emph{optimal completion time}, the earliest possible time that the 
download could complete. This is the time that the seed has uploaded a complete copy 
of the content. For this setup, this time is around 650 seconds into the
experiment.

Fast leechers complete their download soon after the optimal
completion time.  Medium and, especially, slow leechers take
significantly longer to finish.  Thus, contributing to the torrent
enables a leecher to enter the fast cluster and receive data at higher
rates. This in turn ensures a short download completion time. The
choking algorithm does indeed foster reciprocation by rewarding
contributing peers.  In experiments with upload limits following a
uniform distribution, the peer completion time is also uniform;
completion time decreases when a peer's upload contribution
increases.  This further indicates the algorithm's consistent
properties with respect to maintaining sharing incentives.

\begin{figure}[t]
\centering
\includegraphics[width=2.9in]
{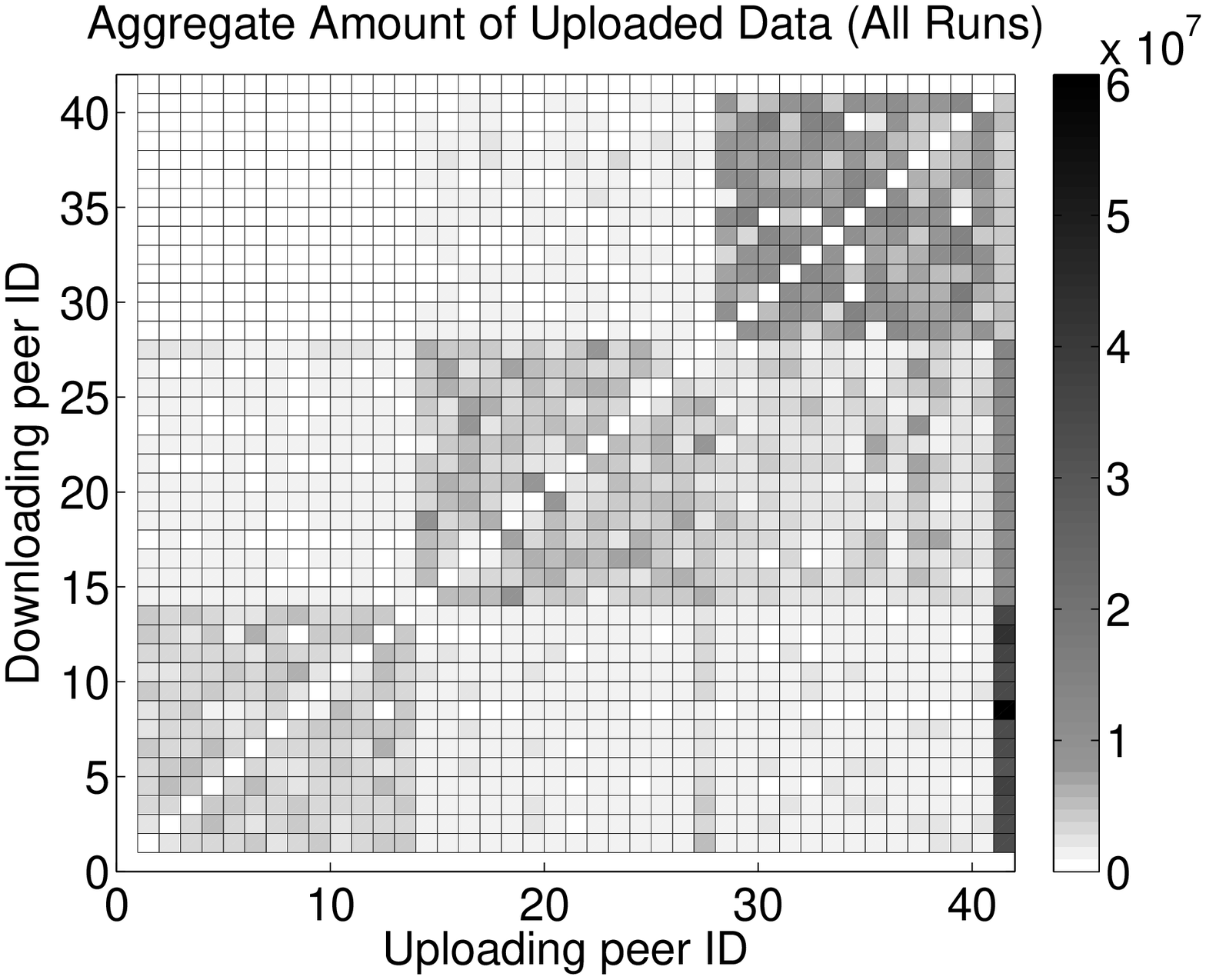}
\caption{\textmd{\textsl{Total number of bytes peers uploaded to each other, 
averaged over all runs. Darker squares represent more data. Peers 1 to 13 
have a 20 kB/s upload limit, peers 14 to 27 have a 50 kB/s upload limit, and peers 28 to 40 
have a 200 kB/s upload limit. The seed (peer 41) is limited to 200 kB/s.
\textit{Fast peers upload much more data than the rest.}}}}
\label{fig:agg-amount-bytes-onethird-seed-200}
\end{figure}

Note, however, that this does not imply any notion of data volume
fairness.  Fast peers end up uploading significantly more data than
the rest.  Figure~\ref{fig:agg-amount-bytes-onethird-seed-200},
which plots the actual volume of uploaded data averaged over all runs,
demonstrates that fast peers are major contributors to the torrent.
Most of their bandwidth is expended on other fast peers, per the
clustering principle. Interestingly, the slow leechers end up
downloading more data from the seed.
The seed provides equal service to peers of any class;
however, slow peers have more opportunity to download from the seed since
they take longer to complete.

In summary, BitTorrent  provides effective incentives for peers to contribute, as 
doing so will reward a leecher with significantly higher download rates. 
Recent studies~\cite{liogkas06, locher06} have shown that limited
free-riding is possible in BitTorrent under specific circumstances,
although such free-riders do not appear to severely impact the quality of
service for other peers. However, these studies do not significantly challenge the
effectiveness of sharing incentives enforced by the choking algorithm. Although 
free-riding is indeed possible, the selfish peers typically achieve lower download rates than
they would if they followed the protocol. As a result, if peers wish to obtain as high
download rates as possible, it is still in their best interest to conform to protocol guidelines.

\begin{figure}[t]
\centering
\includegraphics[width=2.9in]
{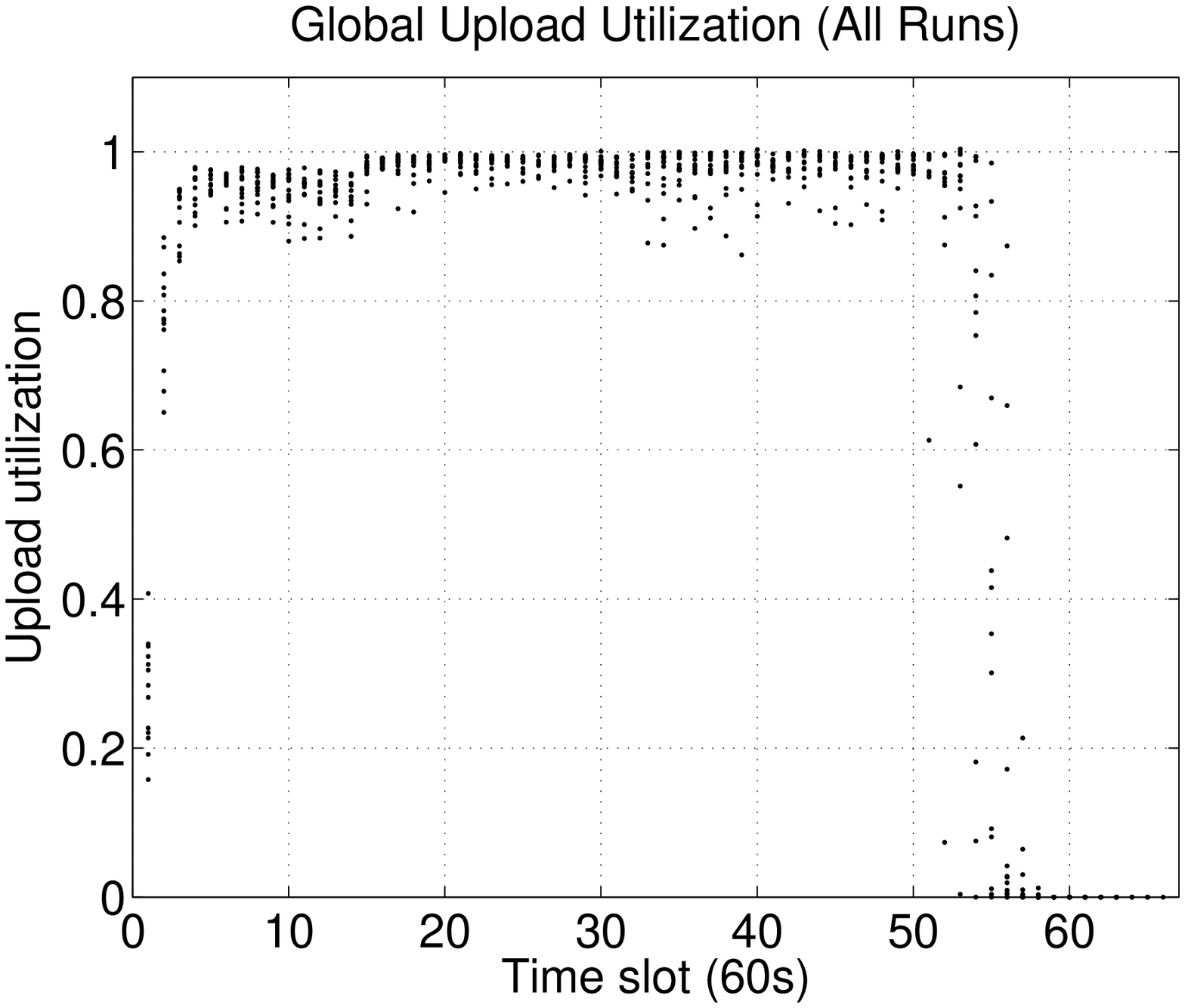}
\caption{\textmd{\textsl{Scatterplot of peers' upload utilization for
      all 60-sec time intervals during the download, in the presence
      of a well-provisioned seed (limited to 200 kB/s).  Each
      dot represents the average upload utilization over all peers for
      a given experiment run.
\textit{Utilization is kept high during most of the download session.}}}}
\label{fig:up-util-cdf-onethird-seed-200}
\end{figure}

\subsubsection{Upload Utilization}
\label{sec:upload-utilization-fast}
We now turn our attention to performance by measuring whether the choking algorithm can
maintain high utilization of the peers' upload capacity. Upload utilization constitutes a 
reliable metric of efficiency in content distribution systems since the total upload 
capacity of all peers represents the maximum throughput the system can achieve as a whole. 
As a result, an efficient protocol should keep peers' upload pipes full at
all times.

Figure~\ref{fig:up-util-cdf-onethird-seed-200} is a scatterplot of peers' upload 
utilization in the aforementioned setup.
A utilization of 1 represents taking full advantage of the available upload 
capacity. 
Utilization for each of the 13 runs is plotted once per minute.  The metric
is torrent-wide: for each minute, we sum the upload bandwidth used by the
peers during that minute, and divide by the upload capacity
available over that minute from all peers still connected at
the minute's end.  The total capacity decreases over time as peers
complete their downloads and disconnect.
Utilization is low at the beginning and the end 
of the session, but close to optimal for the majority of the download.
It increases slightly after approximately 900 seconds, which
corresponds to when fast peers leave the torrent; perhaps the 4-peer limit
on parallel uploads restricts fast peers' utilization, or perhaps TCP
congestion control's AIMD
dynamics have more impact at higher bandwidths.
Nevertheless, utilization is good overall.

In summary, the choking algorithm, in cooperation with 
other BitTorrent mechanisms such as rarest-first piece selection, does a good job
of ensuring high utilization of the upload capacity of leechers
during most of the download. 
We discuss a potential solution to low upload utilization at the beginning
of a leecher's download in Section~\ref{sec:track-prot-extens}.

\subsubsection{Seed Service}

\begin{figure}[t]
\centering
\includegraphics[width=2.9in]{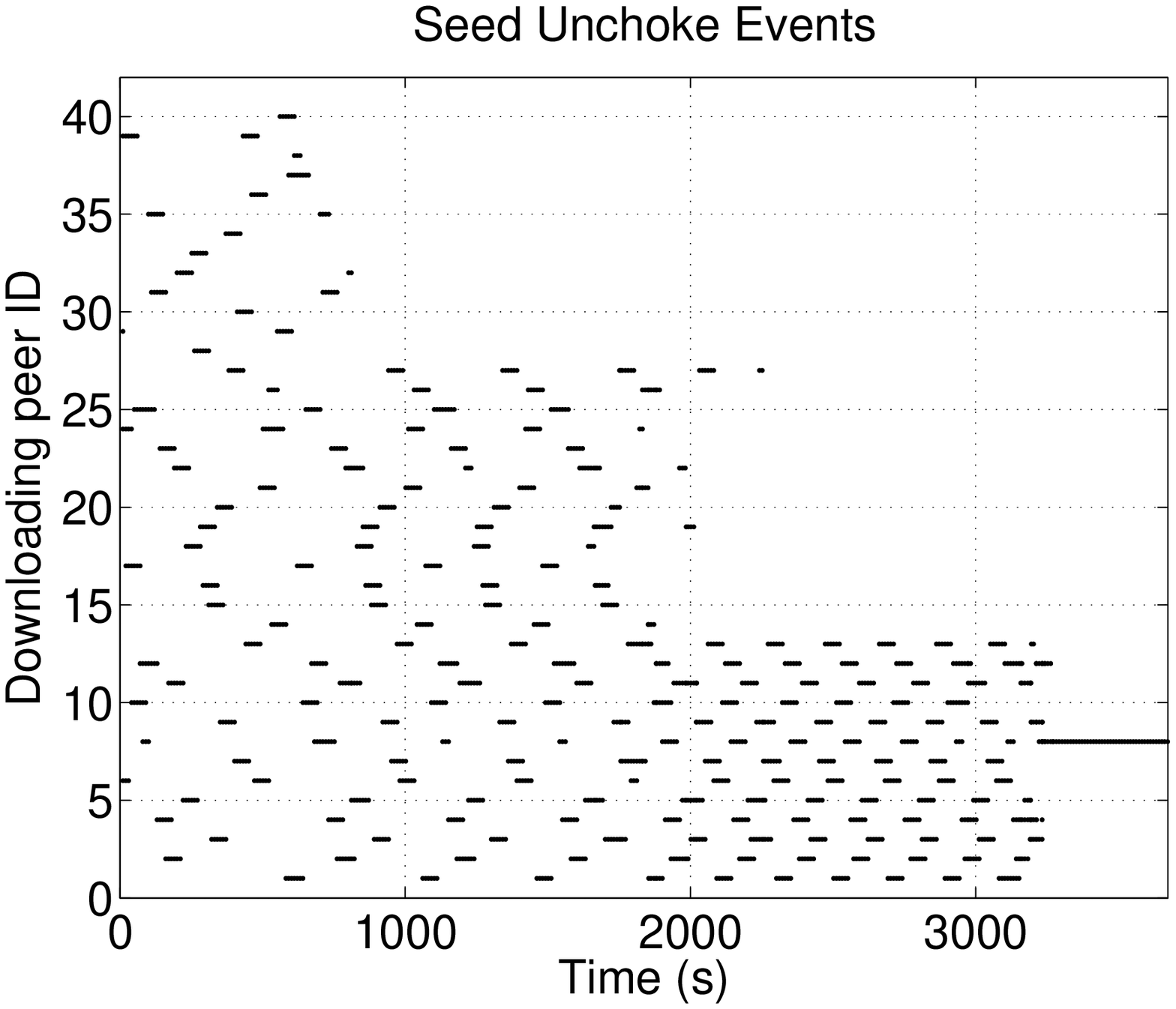}
\caption{\textmd{\textsl{Duration of all unchokes (regular and optimistic) performed by the 
seed to each peer. Results for a single representative run. 
Peers 1 to 13 have a 20 kB/s upload limit, peers 14 to 27 have a 50 kB/s upload limit, and 
peers 28 to 40 have a 200 kB/s upload limit. The seed (peer 41) is limited to 200 kB/s. 
\textit{The seed provides uniform service to all leechers.}}}}
\label{fig:seed-sku-sru-seed-200}
\end{figure}

The official BitTorrent client's choking algorithm for seeds changed as of
version 4.0.0, as described in Section~\ref{choking_algorithm}.
The client's version notes claim that this new algorithm ``addresses the problem for 
which super-seeding was created, but without its problems''. 
We performed detailed 
experiments to study this new algorithm for the first time, and examine this claim. 

Figure~\ref{fig:seed-sku-sru-seed-200} shows the duration of unchokes, both regular and
optimistic, performed by the seed in a representative run of the aforementioned setup. 
Leechers are unchoked in a uniform manner, regardless of upload speed. 
Fast peers, those with higher peer IDs, complete their 
download sooner, after which time the seed divides its upload bandwidth among the remaining 
leechers. Leecher 8 is the last to complete (as shown in
Figure~\ref{fig:download-speed-onethird-seed-200}), and receives exclusive service from 
the seed during the end of its download. 
We see that the new choking algorithm in seed state provides uniform service; this is because
it takes each leecher's waiting time into account. As a result, the risk 
of fast leechers downloading the entire content and quickly disconnecting from the
torrent is reduced. Furthermore, this behavior might help mitigate the effectiveness of 
exploits that attempt to monopolize the seeds~\cite{liogkas06}.

\begin{figure}[t]
\centering
\includegraphics[width=2.9in]
{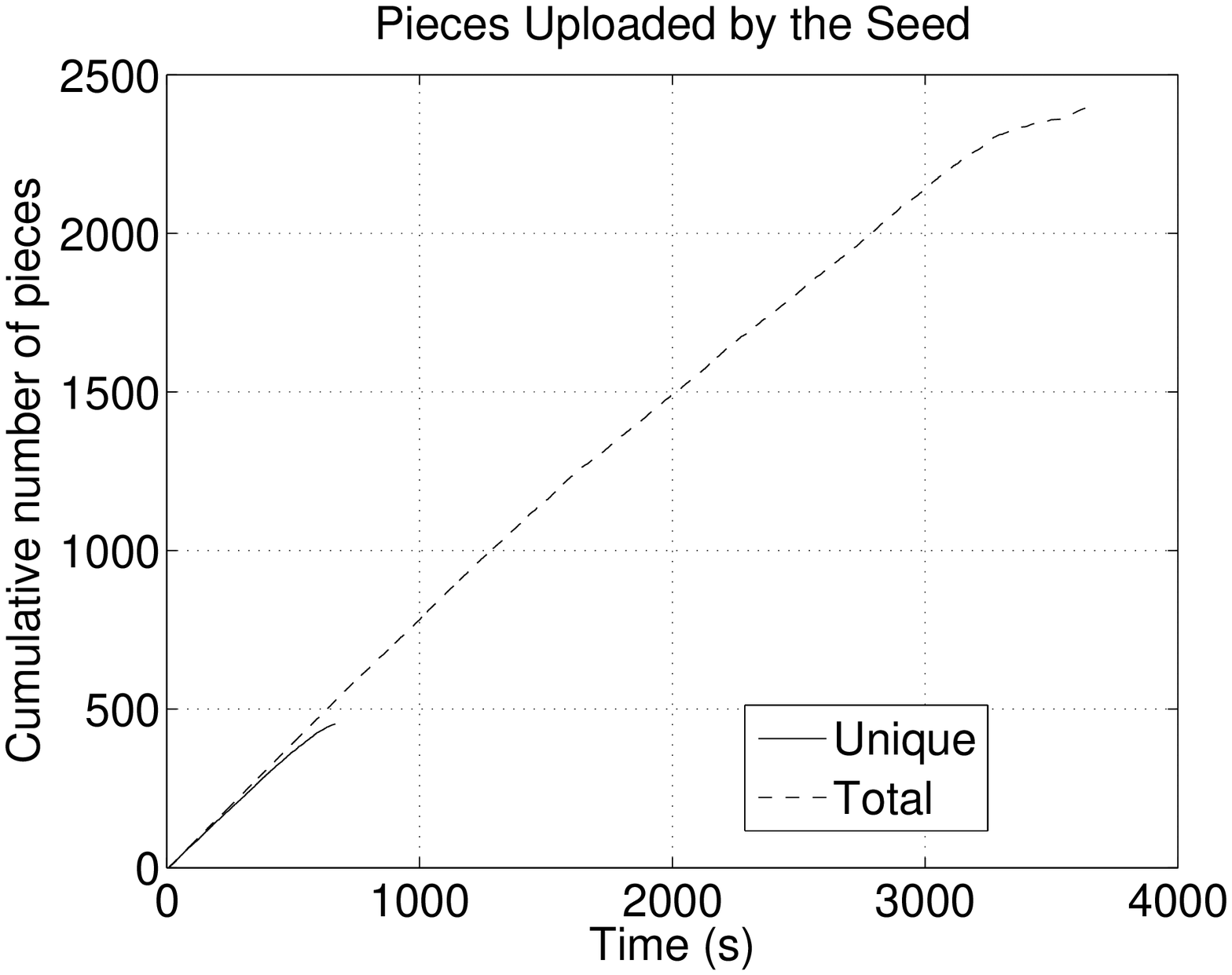}
\caption{\textmd{\textsl{Number of pieces uploaded by the seed, for a single representative run. 
The Unique line represents the pieces that had not been previously uploaded, 
while the Total line represents the total number of pieces uploaded so far.
The seed is limited to 200 kB/s. 
\textit{We observe a 16\% duplicate piece overhead.}}}}
\label{fig:seed-unique-piece-seed-200}
\end{figure}

According to anecdotal evidence~\cite{btwikispec}, seeds using the
pre-4.0.0 choking algorithm might have to upload 150\% to 200\% of the 
total content size before other peers became seeds. In our experiments, the 
new choking algorithm avoids this problem.
Figure~\ref{fig:seed-unique-piece-seed-200} plots the number of pieces uploaded by the seed
during the download session for a representative run. 
527 pieces are sent out before an entire copy of the content
(453 pieces) has been uploaded. Thus, the duplicate piece overhead is around 16\%, 
indicating that the new choking algorithm in seed state avoids
unnecessarily uploading duplicate pieces to a certain extent. 
This number was consistent across all our experiments. 
However, to the best of our knowledge, there has been no experimental evaluation 
of the corresponding overhead in the old choking algorithm in seed state, so it is not clear how 
much of an improvement this is; we will investigate this in future work.

Nevertheless, 16\% duplication represents an opportunity for improvement.
The existing
implementation always issues requests for pieces in the rarest pieces set in the same order, if the set
contains more than one. As a result, leechers might end up requesting the same rarest 
piece from the seed at approximately the same time. It would arguably be preferable for 
leechers to request rarest pieces in random order, so that the probability of multiple
leechers requesting the same piece at the same time is minimized.

\begin{figure}[t]
\centering
\includegraphics[width=2.9in]
{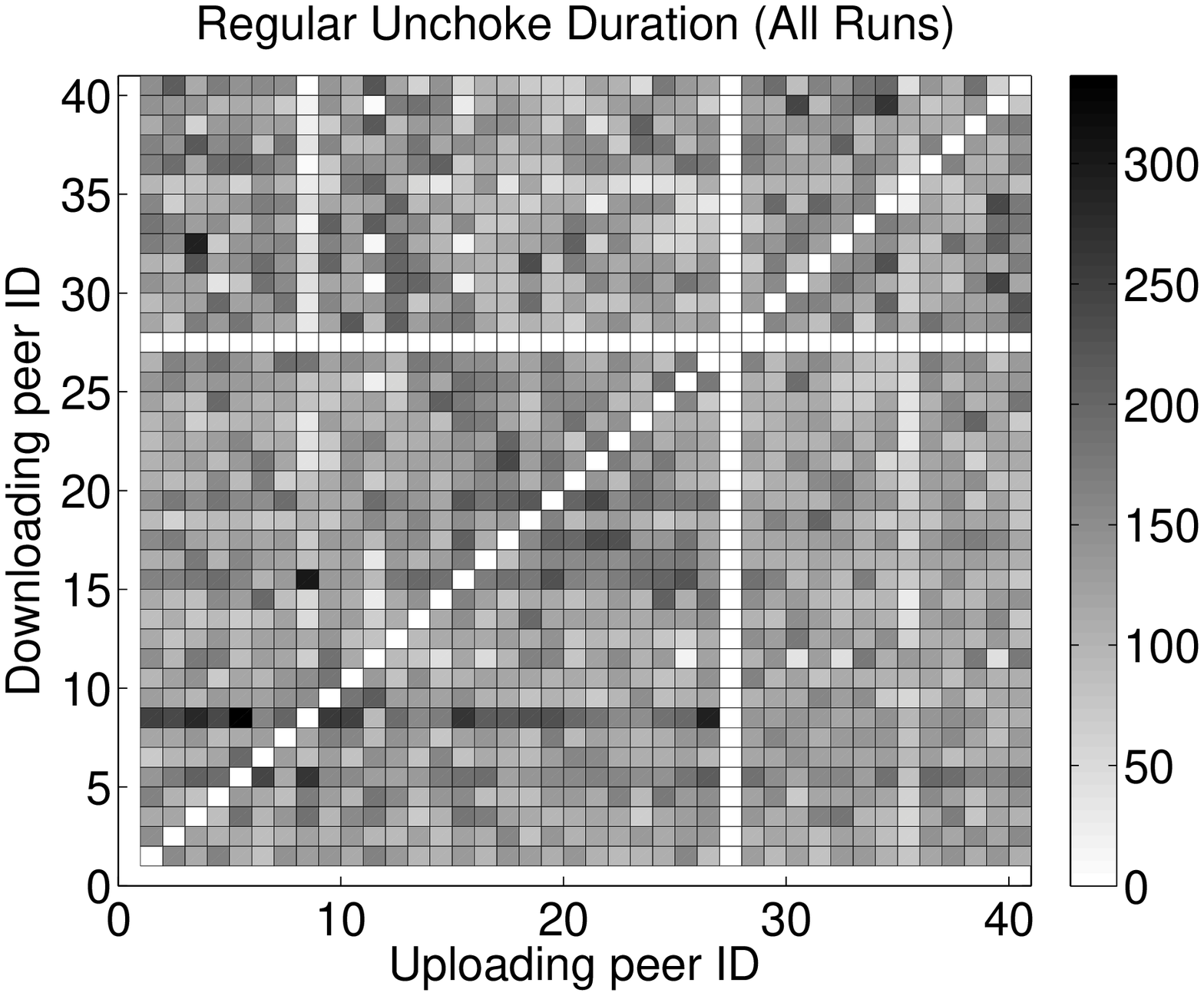}
\caption{\textmd{\textsl{Time duration that peers unchoked each other via a regular unchoke, 
averaged over all runs. Darker squares represent longer unchoke times. 
Peers 1 to 12 have a 20 kB/s upload limit, peers 13 to 26 have a 50 kB/s upload limit, 
and peers 28 to 40 have a 200 kB/s upload limit. The seed (peer 27) is limited to 100 kB/s.
\textit{There is no discernible clustering.}}}}
\label{fig:corr-unchoke-upload-onethird-seed-medium}
\end{figure}

In summary, the new choking algorithm in seed state uniformly
distributes seed upload capacity among leechers, independently of 
their upload contributions.  Our results also show that it
incurs a reasonably low duplicate piece overhead. 

\subsection{Underprovisioned Initial Seed}
\label{under_provisioned_seed}

\begin{figure}[t]
\centering
\includegraphics[width=2.9in]
{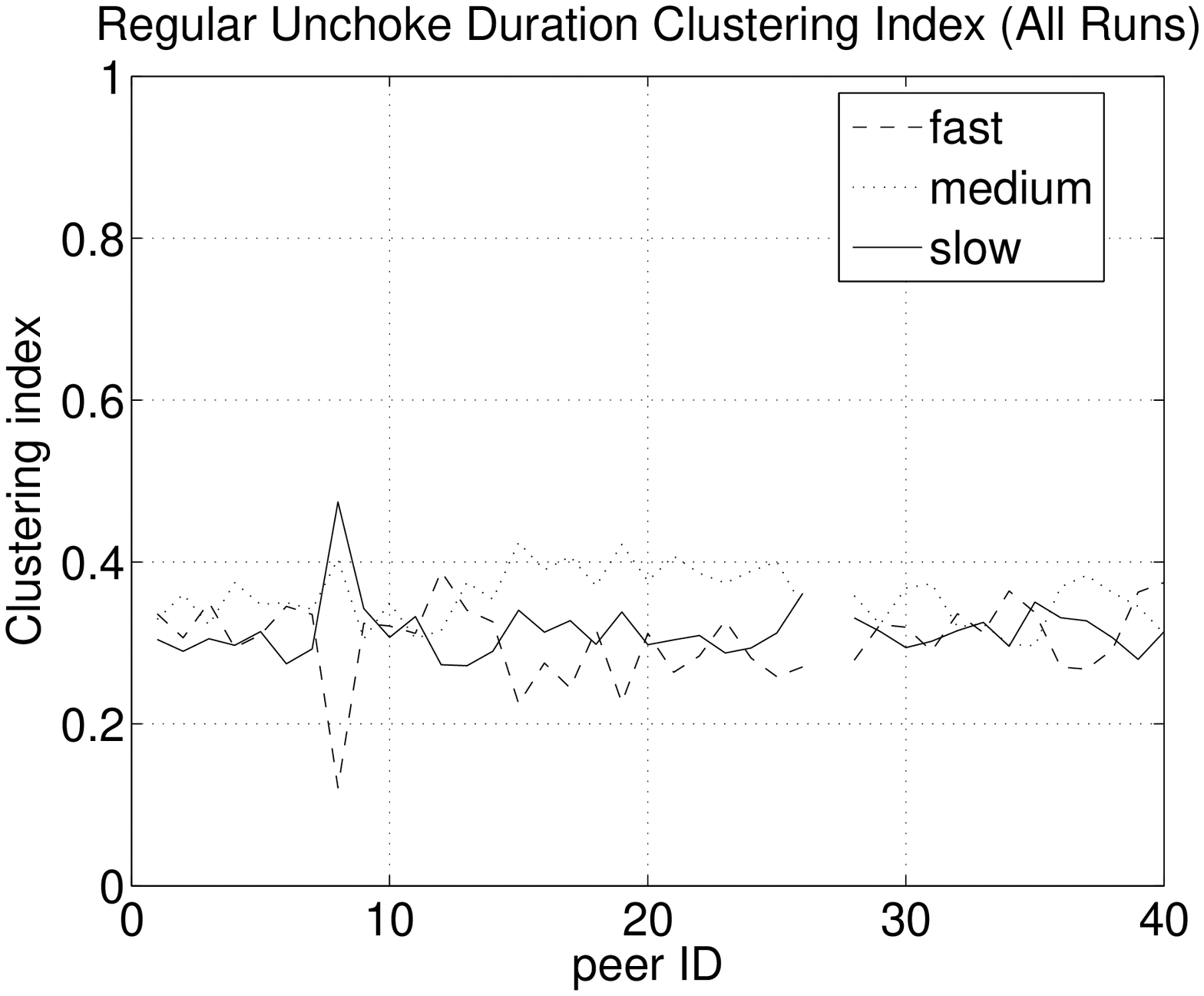}
\caption{\textmd{\textsl{Clustering indices for all peers in the presence of an underprovisioned seed. 
Peers 1 to 12 have a 20 kB/s upload limit, peers 13 to 26 have a 50 kB/s upload limit, 
and peers 28 to 40 have a 200 kB/s upload limit. The seed (peer 27) is limited to 100 kB/s.
\textit{Peers do not show a clear preference to unchoke other peers in any particular class.}}}}
\label{fig:clustering-index-onethird-seed-medium}
\end{figure}

We now turn our attention to a scenario with an underprovisioned
initial seed and demonstrate that the seed upload capacity is
critical to performance during the beginning of a torrent's lifetime.
The experiment we present here involves a single seed and 39 leechers,
12 slow, 14 medium, and 12 fast.  The initial seed in this case, represented
as peer 27 in the following figures, is limited to 100~kB/s, not
200~kB/s. We set the number of parallel uploads again to 4 for the seed and
all the leechers. The results we present are based on 8 experiment
runs, and are consistent with our observations for experiments with
other torrent configurations.  We show that peer behavior in the
presence of an underprovisioned initial seed is substantially
different than with a well-provisioned initial seed.
 
\subsubsection{Clustering}
\label{sec:clustering-slow}

\begin{figure}[t]
\centering
\includegraphics[width=2.9in]
{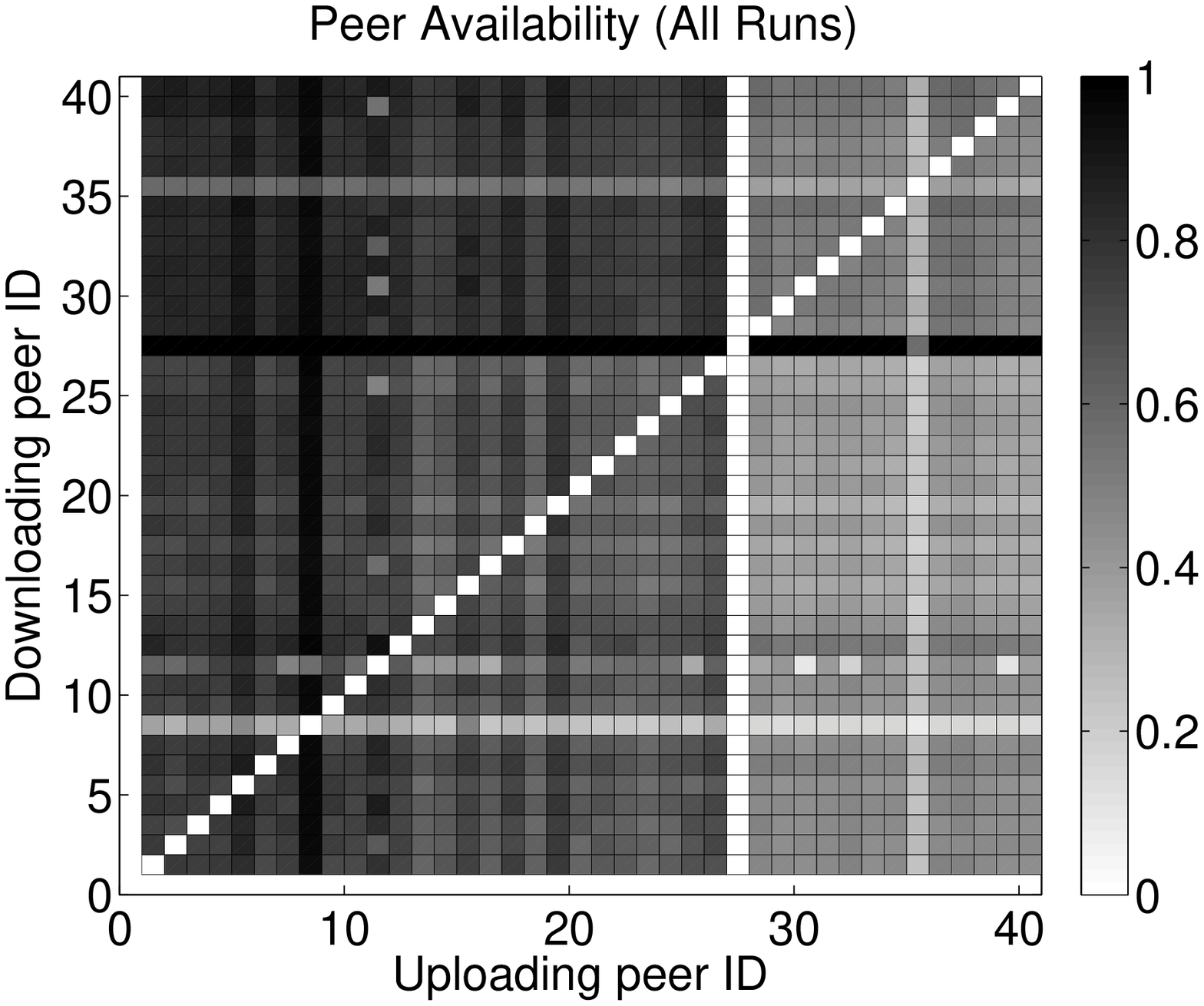}
\caption{\textmd{\textsl{Normalized interested time duration for each
      peer pair, averaged over all runs. Darker squares represent
      higher peer availability.  Peers 1 to 12 have a 20 kB/s upload
      limit, peers 13 to 26 have a 50 kB/s upload limit, and peers 28
      to 40 have a 200 kB/s upload limit. The seed (peer 27) is
      limited to 100 kB/s.  \textit{Fast peers have poor peer availability
        to all other peers.}}}}
\label{fig:cumul-interest-run1-seed-medium}
\end{figure}

Figure~\ref{fig:corr-unchoke-upload-onethird-seed-medium} shows the total time peers 
unchoked each other via a regular unchoke, averaged over all runs of the experiment. 
In contrast to Figure~\ref{fig:corr-unchoke-upload-onethird-seed-200}, 
there is no discernible clustering among peers in the same class. 
The lack of clustering in the presence of an underprovisioned initial seed becomes more apparent
when considering the clustering index metric mentioned in Section~\ref{clustering-fast}.
Figure~\ref{fig:clustering-index-onethird-seed-medium} shows the clustering indices of 
all peers. They are all very similar, indicating a lack of preference to unchoke peers
in any particular class. Compare this to Figure~\ref{fig:clustering-index-onethird-seed-fast}, 
where the preference for peers in the same class is evident.

Figure~\ref{fig:cumul-interest-run1-seed-medium} explains this
behavior by plotting the \textit{peer availability} of each peer
 to each other peer, averaged over all runs of the experiment. We define the \emph{peer availability}
of a downloading peer $Y$ to an uploading peer $X$ as the ratio
of the time $X$ was interested in $Y$ to the time that $Y$
spent in the peer set of $X$.  A peer availability of 1 means that
the uploading peer was always interested in the downloading peer, while a
peer availability of 0 means that the uploading peer was never
interested in the downloading peer.

From the figure we can see that the fast peers have poor peer
availability to all other peers. The seed is uploading new
pieces at a low rate, so even if the seed uploaded only to fast peers,
those fast peers would quickly replicate every piece as it was
completed, remaining idle for the rest of the time. Thus, fast peers are
not interested in others most of the time. The same is not
true for slow peers, since they upload even more slowly than the seed.
In addition, when a fast leecher is unchoked by a slow leecher, it will always
reciprocate with high rates, and thereby be preferred by the slow
leecher. As a result, fast peers will get new pieces even from
medium and slow peers.
Thus, fast peers prevent clustering by
taking up slower peers' unchoke slots and thus breaking any clusters
that might be starting to form. Further experiments with other torrent
configurations, including one with the initial seed further limited to
20~kB/s, confirm this conclusion.

In summary, when the initial seed is underprovisioned, the choking algorithm
does not enable peer clustering. We study in the next section how
this lack of clustering affects the effectiveness of sharing incentives.

\subsubsection{Sharing Incentives}
\label{sec:incentives-slow}
\begin{figure}[t]
\centering
\includegraphics[width=2.9in]
{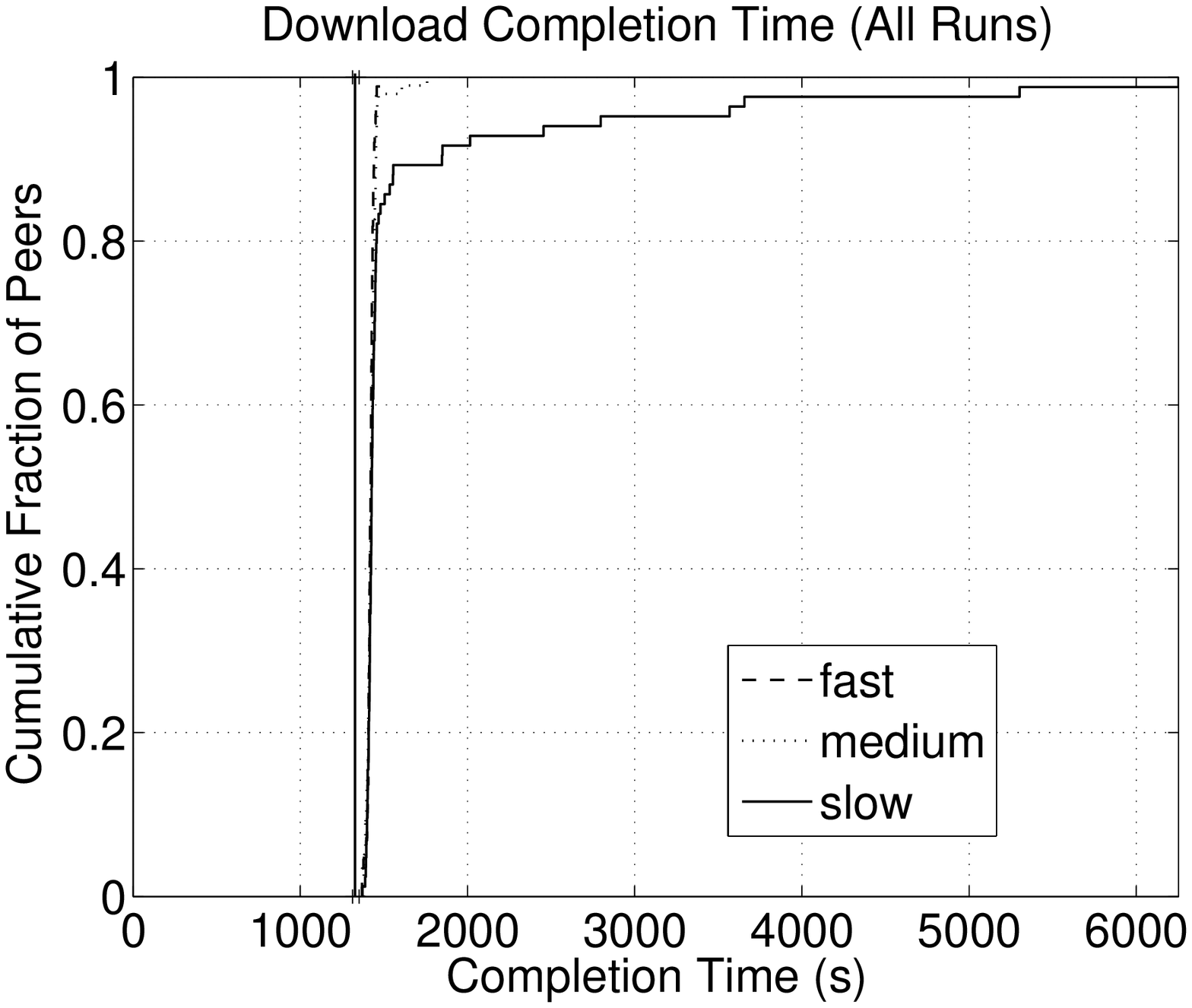}
\caption{\textmd{\textsl{Cumulative distribution of the download completion time for 
the three different classes of leechers, in the presence of an
underprovisioned seed (limited to 100 kB/s), for all runs. 
The vertical line represents the earliest possible time that the download could complete.  \textit{Most peers complete at approximately the same time, soon after the seed finishes uploading a full copy of the content.}}}}
\label{fig:completion-cdf-onethird-seed-medium-class}
\end{figure}

Given the lack of clustering, we now examine whether BitTorrent's 
choking algorithm still provides incentives to share even in the presence 
of an underprovisioned initial seed. In particular, we examine whether fast peers still 
complete their download sooner than others.
Figure~\ref{fig:completion-cdf-onethird-seed-medium-class} shows that this is no 
longer the case.  Most peers complete their download at approximately the same time.
Most points in the tail of the figure
are due to a
single slow peer, peer 8, which in every run completed its download last.
This PlanetLab node has a poor
effective download speed independently of the choking algorithm, 
likely due to network problems or machine overload.
All other peers achieve completion times below 2000 seconds in every
experiment.
Clearly, seed upload capacity is the performance bottleneck.
Once the seed finishes uploading a full copy of the content, all peers complete soon 
thereafter.
Since uploading data to other peers does not shorten a peer's completion
time, BitTorrent's sharing incentives are ineffective here.

\begin{figure}[t]
\centering
\includegraphics[width=2.9in]
{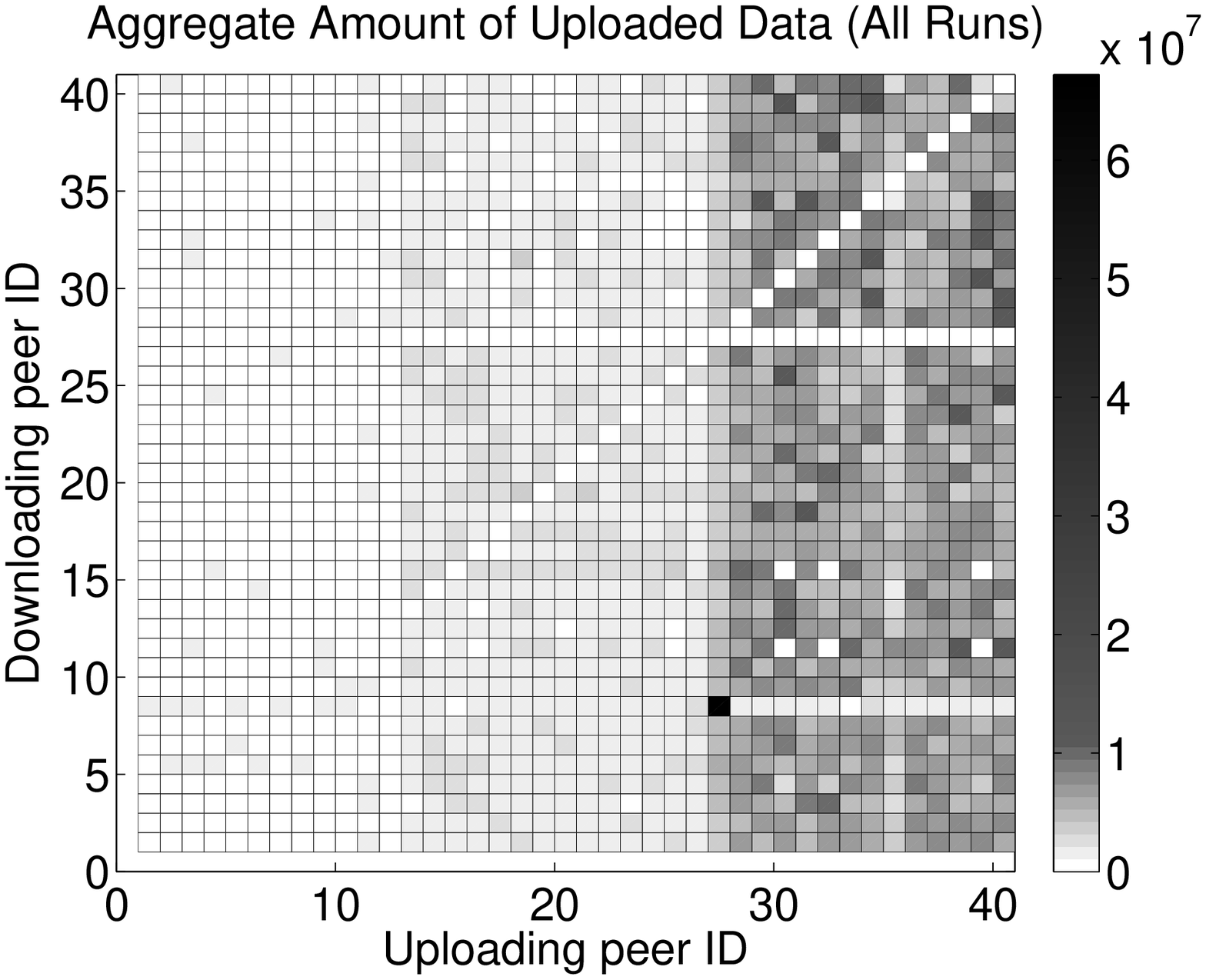}
\caption{\textmd{\textsl{Total number of bytes peers uploaded to each other, 
averaged over all runs. Darker squares represent more data. 
Peers 1 to 12 have a 20 kB/s upload limit, peers 13 to 26 have a 50 kB/s upload limit, 
and peers 28 to 40 have a 200 kB/s upload limit. The seed (peer 27) is
limited to 100 kB/s. \textit{Fast peers upload much more data than the rest,
distributing those data evenly among all peers.}}}}
\label{fig:agg-amount-bytes-onethird-seed-medium}
\end{figure}

Fast peers are again the major contributors in the torrent, but in this
case their upload bandwidth is expended equally across other fast peers and
slower peers alike.  Figure~\ref{fig:agg-amount-bytes-onethird-seed-medium}
plots the amount of uploaded data between each peer pair. A quick visual
inspection shows that the fast peers contribute roughly equally to
all other peers, and that fast peers made most
contributions, while the slow ones made the least.

In summary, when the initial seed is underprovisioned, the choking algorithm
does not provide effective incentives to share. However, the available
upload capacity of fast peers is effectively utilized to
efficiently replicate the pieces being sent by the initial seed.

\subsubsection{Upload Utilization}
\label{sec:upload-util-slow}
\begin{figure}[t]
\centering
\includegraphics[width=2.9in]
{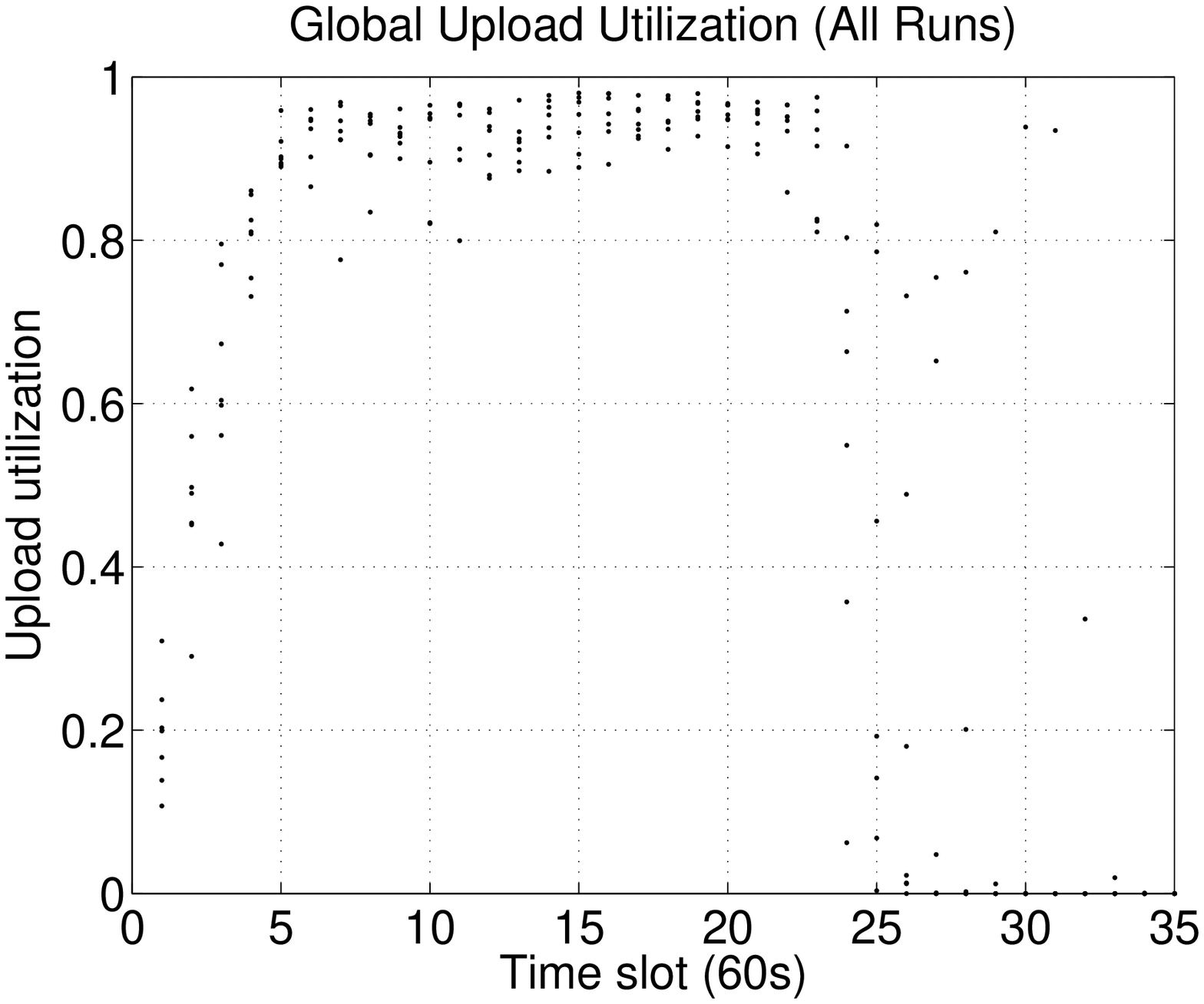}
\caption{\textmd{\textsl{Scatterplot of peers' upload utilization for all 60-sec time 
intervals during the download, in the presence of an underprovisioned
seed (limited to 100 kB/s). 
Each dot represents the average upload utilization over all peers for a given experiment run. 
\textit{Utilization is kept at acceptable levels despite the seed limitation.}}}}
\label{fig:up-util-cdf-onethird-seed-medium}
\end{figure}

We now evaluate the impact of an underprovisioned initial seed on overall \bt system
performance.  Figure~\ref{fig:up-util-cdf-onethird-seed-medium} plots
peers' upload utilization.  Even with a slow seed, upload utilization remains relatively high.
Leechers manage to exchange data productively among themselves once new pieces are downloaded 
from the slow seed, so that the lack of clustering does not degrade torrent performance significantly. 
Interestingly, the BitTorrent design seems to lead the system to do the right thing: fast peers 
contribute their bandwidth to reduce the burden on the initial seed, helping disseminate the available
pieces to slower peers. Indeed, this destroys clustering, but it
improves the torrent efficiency, which is a reasonable decision given the situation.

\begin{figure}[t]
\centering
\includegraphics[width=2.9in]
{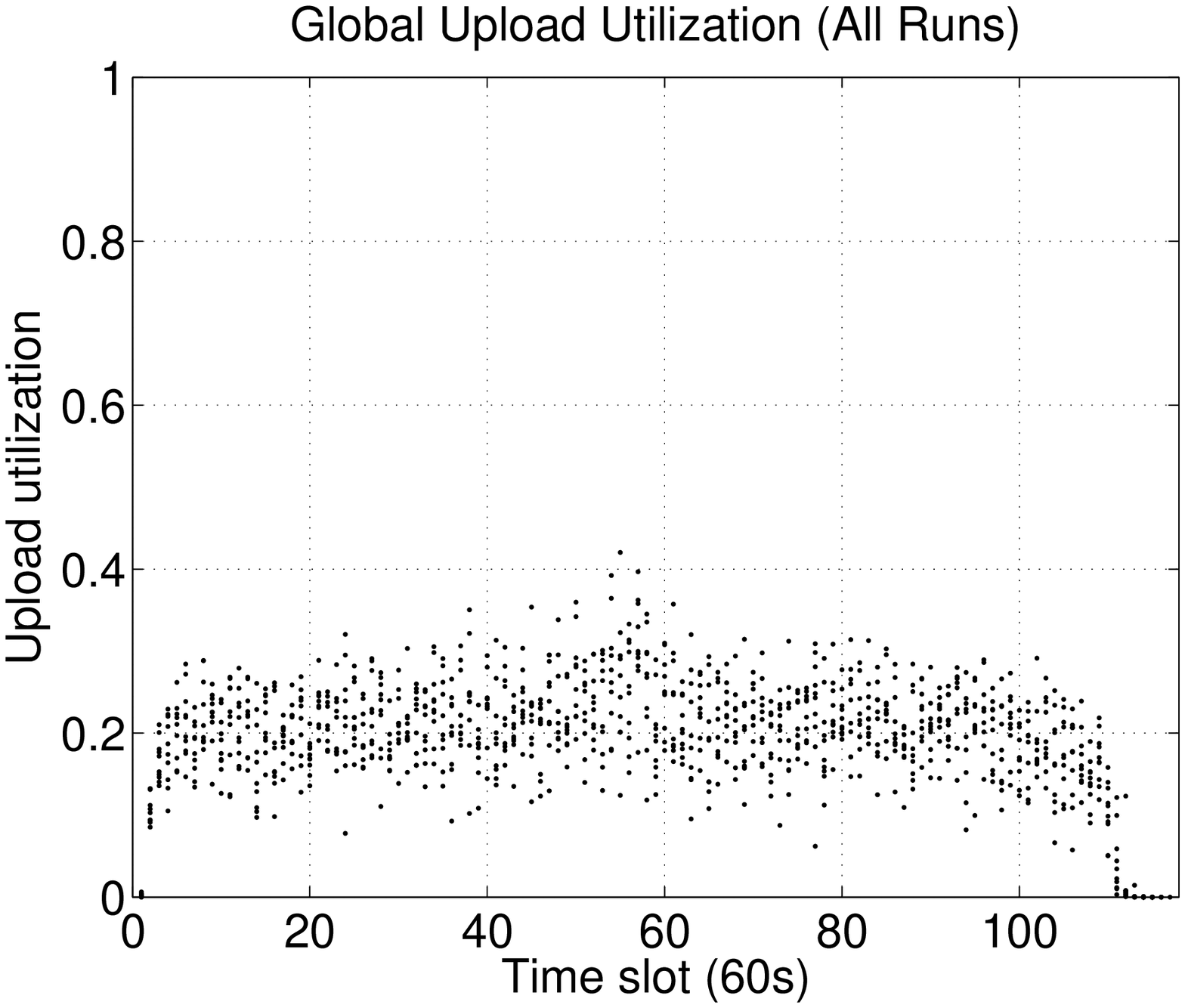}
\caption{\textmd{\textsl{Scatterplot of peers' upload utilization for all 60-sec time 
intervals during the download, in the presence of a severely underprovisioned
seed (limited to 20 kB/s). 
Each dot represents the average upload utilization over all peers for a given experiment run. 
\textit{Utilization is poor when the seed is very slow.}}}}
\label{fig:up-util-cdf-onethird-seed-slow}
\end{figure}

We also experimented with a seed limited to an upload capacity of 20~kB/s. With
this extremely low seed upload speed, there are few new pieces available to
exchange at any point in time, and each new piece gets disseminated rapidly
after it is retrieved from the seed.
Fig.~\ref{fig:up-util-cdf-onethird-seed-slow} shows that the overall upload utilization is
now low; slow peers exhibit slightly higher utilization than the
rest, since they do not need many available pieces to use up their
available upload capacity. 

In summary, even in situations where the initial seed is underprovisioned, the global
upload utilization can be high. However, our
experiments only involve collaborative users, who do not
try to adapt their upload speed according to a utility function of the
observed download speed. On the other hand, in a selfish environment with an
underprovisioned seed, one might expect  a
lower upload utilization due to the lack of sharing incentives. 

\section{Discussion}
\label{discussion}
We discuss two limitations of the choking
algorithm that we identified in the section~\ref{results}: seed upload
capacity is fundamental to the proper operation of the incentives
mechanism, and at the beginning of the download session peers take
some time to reach full upload utilization.

\subsection{Seed Provisioning}
\label{sec:seed-provisioning}
When the initial seed is underprovisioned, the choking algorithm does not
lead to clustering of similar-bandwidth peers. Even 
without clustering, however, we observed high upload utilization.
Interestingly, in the presence of a slow initial seed, the protocol makes fast leechers 
contribute to the download of all other peers, fast or slow, as evidenced in 
Figure~\ref{fig:agg-amount-bytes-onethird-seed-medium}, thereby improving 
the overall torrent capacity. 

However, whenever feasible, one should engineer adequate initial seed capacity 
in order to allow fast leechers to achieve high performance. Our results show that 
the lack of clustering occurs when fast peers cannot maintain their interest in other
fast peers. In order to avoid this situation, the initial seed should at least be able to 
upload data at a speed that matches that of the fastest peers in the torrent.
This suggestion is simply a rule-of-thumb guideline, and assumes that the service
provider knows a priori the maximum upload capacity of the peers that may
join the torrent in the future.
In practice, reasonable bounds could be derived from measurements or from an
analysis of deployed network technologies.
Further research is needed to evaluate the exact impact of seed capacity.
We are currently developing an analytical model that attempts to
express the effect of initial seed capacity on the overall torrent performance.

\subsection{Tracker Protocol Extension}
\label{sec:track-prot-extens}
When a new leecher first joins the torrent, it connects to a random subset 
of already-connected peers that are returned by the tracker. However, in order to
reach its optimal bandwidth utilization, this new peer needs
to exchange data with those peers that have a similar upload capacity 
to itself. If there are few such peers in the 
torrent, it may take some time to discover them, since this process has to be
done via random optimistic unchokes that take place only once every 30
seconds.

Consequently, it might be preferable in such a scenario to employ the 
tracker to assist in matching similar-bandwidth leechers. In this manner,
the discovery period duration could decrease and the upload utilization would
be high even at the beginning of a peer's download.
The new peer could report its upload capacity to the tracker when joining
the torrent. This speed can be the one configured in the 
client software, or possibly the actual maximum upload speed measured 
during previous downloads. The tracker would then reply with a \emph{random subset} 
of peers as usual, along with their upload capacity. The new leecher would 
have the option of performing optimistic unchokes first to peers with 
upload capacity similar to its own, in an effort to discover the best partners
sooner.

With this new tracker protocol extension, if the peer set contains only a few
leechers with similar upload capacity, they will be discovered quickly.
However, since the tracker still returns a random subset of peers independently 
of the advertised upload capacity, there is no benefit for a peer to lie. 
If it does so, other peers who connect to it will discover this fact quickly, and
choke the lying leecher, since it would not be able to sustain appropriate upload rates. In a 
collaborative environment, however, the tracker might even want to return peers based on 
their advertised upload capacity, as also proposed in~\cite{bharambe06}, in order to 
speed up cluster creation even more. 
Although this extension is promising, further research is required to verify that 
it will work as expected. 

\section{Related Work}
\label{related}
There has been a fair amount of work on the performance and behavior of 
BitTorrent systems.
Bram Cohen, the protocol's creator, has described BitTorrent's main 
mechanisms and their design rationales~\cite{cohen03}. 
Several analytical studies have formulated models for BitTorrent-like 
protocols.
Biersack \textit{et al.}~\cite{biersack04} propose an analysis of
three content distribution models: a linear chain, a tree, and a
forest of trees. They discuss the impact of the number of pieces and
the number of parallel uploads for each model, and claim that the
optimal efficiency is achieved using 3 to 5 parallel uploads.  
Yang \textit{et al.}~\cite{yang04} study the service capacity of 
BitTorrent systems and show that it increases exponentially at the 
beginning of the torrent, and scales well with the number of peers.  
Qiu \textit{et al.}~\cite{qiu04} extend this work by providing an 
analytical solution to a fluid model of BitTorrent.  Their results show 
BitTorrent's high upload utilization. However, their model assumes peer
selection based on global knowledge of all peers in the torrent, as well
as uniform distribution of pieces. Moreover, they do not consider the
dynamics of the choking algorithm. 
Massoulie \textit{et al.}~\cite{massoulie06} introduce a probabilistic 
model and claim that system performance does not depend critically on 
the rarest-first piece selection strategy.  
Lastly, Fan \textit{et al.}~\cite{fan06} characterize the complete design
space of BitTorrent-like protocols by providing a mathematical model
that captures the trade-off between high performance and fairness.
As previously mentioned, whereas all these models provide valuable insight 
into the behavior of BitTorrent systems, unrealistic assumptions limit their 
applicability in real scenarios.

Other researchers have relied on simulations to understand BitTorrent's properties.
Felber \textit{et al.}~\cite{felber04} compare different peer and piece
selection strategies in different torrent configurations.
Bharambe \textit{et al.}~\cite{bharambe06} utilize a discrete event simulator to
evaluate upload utilization and bit-level fairness. They find that the protocol 
scales very well and that the rarest-first algorithm outperforms alternative piece 
selection strategies. However, they do not evaluate a peer set larger than 15 peers,
whereas the official implementation has a default value of 80. This limitation
may have an important impact on the behavior of the protocol, as the accuracy of the
piece selection strategy is affected by the peer set size. Moreover, they do
not consider the new version of the choking algorithm in seed state. 
Tian \textit{et al.}~\cite{tian06} propose a simple analytical model to study
BitTorrent's performance and validate it using simulations. They also propose and 
evaluate a new peer selection strategy during the last phase of a download session, 
in order to enable more peers to complete their download after the departure of all 
the seeds.

There have been several measurement studies that examined actual BitTorrent traffic.
Izal \textit{et al.}~\cite{izal04} identify several peer characteristics in the tracker 
log for the Redhat Linux 9 ISO image, including the percentage of peers completing 
the download, load on the seeds, and geographical spread of participating peers.
They observe a correlation between uploaded and downloaded amount of data.
Pouwelse \textit{et al.}~\cite{pouwelse05} study the file popularity, file
availability, download performance, and content lifetime on a formerly popular 
tracker website. They observe that, although BitTorrent can efficiently handle large 
flash crowds, the central tracker component could potentially be a bottleneck.
A more recent study by Guo \textit{et al.}~\cite{guo05} demonstrates that 
peer performance fluctuates widely in small torrents, and that high-bandwidth 
peers tend to contribute less to the torrents. Inter-torrent collaboration is 
proposed as an alternative to providing extra incentives for leechers to stay 
connected after the completion of their download.
Lastly, Legout \textit{et al.}~\cite{legout06} run extensive
experiments on real torrents, from the viewpoint of a single peer.
They show that the rarest-first and choking algorithms play a critical
role in BitTorrent's performance. In particular, they show that the
rarest-first piece selection strategy approximates an optimal piece selection strategy
after a complete copy of the content has been uploaded, and that the choking algorithm 
fosters reciprocation. They claim that the replacement of the current choking 
algorithm by a bit-level tit-for-tat algorithm is not appropriate, as proposed 
by other researchers~\cite{jun05}. However, they do not identify the reasons 
behind the properties of the choking algorithm, and fail to examine its dynamics 
due to the single peer viewpoint.

Furthermore, researchers have looked into the feasibility of 
circumventing BitTorrent mechanisms to free-ride on the torrent. 
Shneidman \textit{et al.}~\cite{shneidman04} were the first to demonstrate that 
BitTorrent exploits are indeed feasible.
Jun \textit {et al.}~\cite{jun05} argue that the choking algorithm cannot 
prevent free-riding, and propose a new algorithm as a replacement.
Liogkas \textit{et al.}~\cite{liogkas06} design and implement three exploits
that allow a peer that does not contribute to maintain high download rates 
under specific circumstances. However, they show that, even though such peers 
can sometimes obtain more bandwidth, there is no considerable degradation of the 
overall system's quality of service. 
Lastly, Locher \textit{et al.}~\cite{locher06} extend this work by demonstrating 
that limited free-riding is feasible even in the absence of seeds. They also describe 
how free-riding is possible in BitTorrent sharing communities.

Our work differs from all previous studies in its approach and
results. We perform the first extensive experimental study of
BitTorrent in a controlled environment, by monitoring all the peers in
the torrent, and examining the behavior of the \bt system in a variety of
scenarios.  Our results validate protocol properties that have not
been demonstrated experimentally previously, as well as new properties
with respect to the impact of the initial seed on performance.

\section{Conclusion}
\label{conclusion}
In this paper we presented the first experimental investigation of
BitTorrent systems that links per-peer decisions and overall torrent
behavior.  Our results validated
three BitTorrent properties that, though widely believed to hold, have
not been demonstrated experimentally. We showed that the choking
algorithm enables clustering of similar-bandwidth peers,
ensures effective sharing incentives by rewarding peers who contribute with
high download rates, and achieves high upload utilization for the
majority of the download duration. We also examined the properties of
the new choking algorithm in seed state and the impact of initial seed capacity
on the overall \bt system performance. In particular, we showed that an
underprovisioned initial seed does not enable clustering of peers and does
not guarantee effective sharing incentives. However, we showed that even in such a case, the
choking algorithm guarantees an efficient utilization of the available
resources by enforcing fast peers to help other peers with their download. 
Based on our observations, we offered guidelines for content providers
regarding seed provisioning, and discussed a tracker protocol
extension that addresses an identified limitation of the protocol.

This work opens up many avenues for future work. We are currently developing
an analytical model to express the effect of the seed capacity on torrent performance.
It would also be interesting to run experiments with the old choking
algorithm in seed state
and compare its properties to the new algorithm.
In addition, we would like to investigate the impact of different number of 
regular and optimistic unchokes on the protocol's performance and fairness properties. 
It has recently been argued~\cite{fan06} that the trade-off between these two
kind of unchokes is critical.
The current values used by the protocol are intuition-based engineering 
choices; we would like to conduct a systematic evaluation of system 
behavior under different values for these parameters.

\begin{small}

\end{small}


\begin{thebibliography}{10}

\bibitem{btsite}
Bit{T}orrent, {I}nc.
\newblock \url{http://www.bittorrent.com}.

\bibitem{btwikispec}
Bit{T}orrent {S}pecification wiki.
\newblock \url{http://wiki.theory.org/BitTorrentSpecification/}.

\bibitem{btinstrumented}
Instrumented {B}it{T}orrent client.
\newblock
  \url{http://www-sop.inria.fr/planete/Arnaud.Legout/Projects/p2p_cd.html#soft%
ware}.

\bibitem{pssh}
Parallel versions of openssh tools.
\newblock \url{http://www.theether.org/pssh/}.

\bibitem{planetlab}
Planet{L}ab open platform.
\newblock \url{http://www.planet-lab.org}.

\bibitem{bharambe06}
A.~R. Bharambe, C.~Herley, and V.~N. Padmanabhan.
\newblock Analyzing and {I}mproving a {B}it{T}orrent {N}etwork's {P}erformance
  {M}echanisms.
\newblock In {\em Proc. of Infocom'06}, Barcelona, Spain, April 2006.

\bibitem{biersack04}
E.~W. Biersack, P.~Rodriguez, and P.~Felber.
\newblock Performance {A}nalysis of {P}eer-to-{P}eer {N}etworks for {F}ile
  {D}istribution.
\newblock In {\em Proc. of the International Workshop on Quality of future
  Internet Services (QofIS'04)}, Barcelona, Spain, September 29--October 1,
  2004.

\bibitem{cohen03}
B.~Cohen.
\newblock Incentives {B}uild {R}obustness in {B}it{T}orrent.
\newblock In {\em Proc. of the Workshop on Economics of Peer-to- Peer Systems
  (P2PEcon'03)}, Berkeley, CA, June 2003.

\bibitem{fan06}
B.~Fan, D.-M. Chiu, and J.~C. Lui.
\newblock The {D}elicate {T}radeoffs in {B}it{T}orrent-like {F}ile {S}haring
  {P}rotocol {D}esign.
\newblock In {\em Proc. of ICNP'06}, Santa Barbara, CA, November 2006.

\bibitem{felber04}
P.~A. Felber and E.~W. Biersack.
\newblock Self-scaling {N}etworks for {C}ontent {D}istribution.
\newblock In {\em Proc. of the International Workshop on Self-* Properties in
  Complex Information Systems (Self-*'04)}, Bertinoro, Italy, May 31--June 2,
  2004.

\bibitem{guo05}
L.~Guo, S.~Chen, Z.~Xiao, E.~Tan, X.~Ding, and X.~Zhang.
\newblock Measurements, {A}nalysis, and {M}odeling of {B}it{T}orrent-like
  {S}ystems.
\newblock In {\em Proc. of IMC'05}, Berkeley, CA, October 2005.

\bibitem{izal04}
M.~Izal, G.~Urvoy-Keller, E.~W. Biersack, P.~Felber, A.~A. Hamra, and
  L.~Garc\'es-Erice.
\newblock Dissecting {B}it{T}orrent: {F}ive {M}onths in a {T}orrent's
  {L}ifetime.
\newblock In {\em Proc. of PAM'04}, Antibes Juan-les-Pins, France, April 2004.
\newpage
\bibitem{jun05}
S.~Jun and M.~Ahamad.
\newblock Incentives in {B}it{T}orrent {I}nduce {F}ree {R}iding.
\newblock In {\em Proc. of the Workshop on Economics of Peer-to-Peer Systems
  (P2PEcon'05)}, Philadelphia, PA, August 2005.

\bibitem{karagiannis04}
T.~Karagiannis, A.~Broido, N.~Brownlee, kc~claffy, and M.~Faloutsos.
\newblock Is {P2P} dying or just hiding?
\newblock In {\em Proc. of Globecom'04}, Dallas, TX, November 29--December 3,
  2004.

\bibitem{legout06}
A.~Legout, G.~Urvoy-Keller, and P.~Michiardi.
\newblock Rarest {F}irst and {C}hoke {A}lgorithms {A}re {E}nough.
\newblock In {\em Proc. of IMC'06}, Rio de Janeiro, Brazil, October 2006.

\bibitem{liogkas06}
N.~Liogkas, R.~Nelson, E.~Kohler, and L.~Zhang.
\newblock Exploiting {B}ittorrent {F}or {F}un ({B}ut {N}ot {P}rofit).
\newblock In {\em Proc. of IPTPS'06}, Santa Barbara, CA, February 2006.

\bibitem{locher06}
T.~Locher, P.~Moor, S.~Schmid, and R.~Wattenhofer.
\newblock Free {R}iding in {B}it{T}orrent is {C}heap.
\newblock In {\em Proc. of HotNets-V}, Irvine, CA, November 2006.
\newblock To appear.

\bibitem{massoulie06}
L.~Massoulie and M.~Vojnovic.
\newblock Coupon {R}eplication {S}ystems.
\newblock In {\em Proc. of SIGMETRICS'05}, Banff, Canada, June 2005.

\bibitem{pouwelse05}
J.~Pouwelse, P.~Garbacki, D.~Epema, and H.~Sips.
\newblock The {B}it{T}orrent {P2P} file-sharing system: {M}easurements and
  {A}nalysis.
\newblock In {\em Proc. of IPTPS'05}, Ithaca, NY, February 2005.

\bibitem{qiu04}
D.~Qiu and R.~Srikant.
\newblock Modeling and {P}erformance {A}nalysis of {B}it{T}orrent-{L}ike
  {P}eer-to-{P}eer {N}etworks.
\newblock In {\em Proc. of SIGCOMM'04}, Portland, Oregon, August 30--September
  3, 2004.

\bibitem{shneidman04}
J.~Shneidman, D.~Parkes, and L.~Massoulie.
\newblock Faithfulness in {I}nternet {A}lgorithms.
\newblock In {\em Proc. of the Workshop on Practice and Theory of Incentives
  and Game Theory in Networked Systems (PINS'04)}, Portland, OR, September
  2004.

\bibitem{tian06}
Y.~Tian, D.~Wu, and K.~W. Ng.
\newblock Modeling, {A}nalysis and {I}mprovement for {B}it{T}orrent-{L}ike
  {F}ile {S}haring {N}etworks.
\newblock In {\em Proc. of Infocom'06}, Barcelona, Spain, April 2006.

\bibitem{yang04}
X.~Yang and G.~de~Veciana.
\newblock Service {C}apacity of {P}eer to {P}eer {N}etworks.
\newblock In {\em Proc. of Infocom'04}, Hong Kong, China, March 2004.

\end{thebibliography}
\end{document}